\PassOptionsToPackage{unicode}{hyperref}
\PassOptionsToPackage{hyphens}{url}

\documentclass{article}

\usepackage{xcolor}
\usepackage{amsmath,amssymb}
\usepackage{iftex}
\ifPDFTeX
  \usepackage[T1]{fontenc}
  \usepackage[utf8]{inputenc}
  \usepackage{textcomp} 
\else 
  \usepackage{unicode-math} 
  \defaultfontfeatures{Scale=MatchLowercase}
  \defaultfontfeatures[\rmfamily]{Ligatures=TeX,Scale=1}
\fi
\usepackage{lmodern}
\ifPDFTeX\else
\fi
\IfFileExists{upquote.sty}{\usepackage{upquote}}{}
\IfFileExists{microtype.sty}{
  \usepackage[]{microtype}
  \UseMicrotypeSet[protrusion]{basicmath} 
}{}
\makeatletter
\@ifundefined{KOMAClassName}{
  \IfFileExists{parskip.sty}{%
    \usepackage{parskip}
  }{
    \setlength{\parindent}{0pt}
    \setlength{\parskip}{6pt plus 2pt minus 1pt}}
}{
  \KOMAoptions{parskip=half}}
\makeatother
\usepackage{longtable,booktabs,array}
\usepackage{calc} 
\usepackage{etoolbox}
\makeatletter
\patchcmd\longtable{\par}{\if@noskipsec\mbox{}\fi\par}{}{}
\makeatother
\IfFileExists{footnotehyper.sty}{\usepackage{footnotehyper}}{\usepackage{footnote}}
\makesavenoteenv{longtable}
\AtBeginEnvironment{longtable}{\small} 
\AtBeginEnvironment{table}{\small} 

\usepackage{caption}
\usepackage{tabularx}
\usepackage{float}
\usepackage{colortbl} 
\usepackage{doi} 

\usepackage{graphicx}
\makeatletter
\newsavebox\pandoc@box
\newcommand*\pandocbounded[1]{
  \sbox\pandoc@box{#1}%
  \Gscale@div\@tempa{\textheight}{\dimexpr\ht\pandoc@box+\dp\pandoc@box\relax}%
  \Gscale@div\@tempb{\linewidth}{\wd\pandoc@box}%
  \ifdim\@tempb\p@<\@tempa\p@\let\@tempa\@tempb\fi
  \ifdim\@tempa\p@<\p@\scalebox{\@tempa}{\usebox\pandoc@box}%
  \else\usebox{\pandoc@box}%
  \fi%
}
\def\fps@figure{htbp}
\makeatother
\ifLuaTeX
  \usepackage{luacolor}
  \usepackage[soul]{lua-ul}
\else
  \usepackage{soul}
\fi
\usepackage{bookmark}
\IfFileExists{xurl.sty}{\usepackage{xurl}}{} 
\hypersetup{
  pdftitle={The BAD Paradox},
  pdfauthor={Ronald Sielinski},
  hidelinks,
  pdfcreator={LaTeX via pandoc}
}

\title{The BAD Paradox: A Critical Assessment of the Belin/Ambrósio Deviation Model}
\author{Ronald Sielinski}
\date{January 21, 2025}

\begin{document}

\maketitle

\begin{abstract}
The Belin/Ambrósio Deviation (BAD) model is a widely used diagnostic tool for detecting keratoconus and corneal ectasia. The input to the model is a set of z-score normalized $D$ indices that represent physical characteristics of the cornea. Paradoxically, the output of the model, Total Deviation Value ($D_{\text{final}}$), is reported in standard deviations from the mean, but $D_{\text{final}}$ does not behave like a z-score normalized value. Although thresholds like $D_{\text{final}} \ge 1.6$ for "suspicious" and $D_{\text{final}} \ge 3.0$ for "abnormal" are commonly cited, there is little explanation on how to interpret values outside of those thresholds or to understand how they relate to physical characteristics of the cornea. This study explores the reasons for $D_{\text{final}}$'s apparent inconsistency through a meta-analysis of published data and a more detailed statistical analysis of over 1,600 Pentacam exams. The results reveal that systematic bias in the BAD regression model, multicollinearity among predictors, and inconsistencies in normative datasets contribute to the non-zero mean of $D_{\text{final}}$, complicating its clinical interpretation. These findings highlight critical limitations in the model's design and underscore the need for recalibration to enhance its transparency and diagnostic reliability.
\end{abstract}

\section{Introduction}

The Belin/Ambrósio Deviation model (BAD) is widely regarded as an effective solution to screen refractive patients for ectasia and keratoconus \cite{matharu2022}. Multiple studies have demonstrated the model's ability to distinguish between normal and keratoconic eyes \cite{song2023, ding2024, hashemi2016, ramos2012, villavicencio2014, toprak2023, ambrosio2023, ambrosio2017, steinberg2015, shetty2017}.

Despite the consistency of these studies' conclusions, a meta-analysis of their data reveals surprising differences in the statistical parameters of the normal (i.e., non-keratoconic) subgroups in their study populations. In particular, the mean of $D_{\text{final}}$, the output of the BAD regression model, ranges from 0.43 to 1.3.

\begin{table}[htbp]
\caption{Statistical parameters of $D_{\text{final}}$ in published studies.}
\label{tab:d_stats}
\centering
\small
\resizebox{\linewidth}{!}{%
\begin{tabular}{@{}
  >{\centering\arraybackslash}p{0.20\linewidth}
  >{\centering\arraybackslash}p{0.10\linewidth}
  >{\centering\arraybackslash}p{0.20\linewidth}
  >{\centering\arraybackslash}p{0.20\linewidth}
  >{\centering\arraybackslash}p{0.10\linewidth}
  >{\centering\arraybackslash}p{0.10\linewidth}@{}}
\toprule
\textbf{Mean ± SD} & 
\textbf{Median} & 
\textbf{Range} & 
\textbf{Group Size (Normal/ Control)} & 
\textbf{Notes} & 
\textbf{Study [Source]} \\ 
\midrule
	& 0.92 & (0.47, 1.40) & 137 & IQR & \cite{song2023} \\
	1.03 ± 0.58 & & (-0.27, 2.78) & 200 & & \cite{ding2024} \\
	0.96 ± 0.8 & & & 100 & & \cite{hashemi2016} \\
	0.43 ± 0.57 & 0.43 & (-1.20, 2.71) & 200 & & \cite{ramos2012} \\
	0.69 ± 0.58 & & (-1.25, 2.61) & 682 & & \cite{villavicencio2014} \\
	0.81 ± 0.52 & & & 70 & & \cite{toprak2023} \\
	& 0.81 & (-1.13, 2.81) & 1680 & & \cite{ambrosio2023} \\
	0.75 ± 0.56 & 0.80 & ({[}-{]}1.13, 2.35) & 480 &  & \cite{ambrosio2017} \\
	1.3 ± 1.3 & 1.0 & (0.6, 1.6) & 196 & IQR & \cite{steinberg2015} \\
	& 0.98 & (0.62, 1.34) & 42 & IQR & \cite{shetty2017} \\
\bottomrule
\end{tabular}%
}
\end{table}

Based on the reported means, standard deviations, and sample sizes from each study, the difference in means $D_{\text{final}}$ between most of these groups is statistically significant.

Practical significance is more difficult to assess. The \emph{Pentacam User Guide} tells us that $D_{\text{final}}$ has been normalized to its mean value and is reported in standard deviations from the mean \cite{oculus2018}. If $D_{\text{final}}$ is z-score normalized variable, it should have a mean of zero (0). However, the reported mean of $D_{\text{final}}$ for normal subgroups in these studies is invariably \emph{non-zero}---including studies involving Drs. Belin and Ambrósio \cite{ramos2012, ambrosio2023, ambrosio2017}.

These results are paradoxical: How can the mean of a z-score normalized variable deviate by a full standard deviation from itself?

There are a number of possibilities: The individual studies could have normative populations that are systematically different from BAD. Demographic characteristics like patient age \cite{oculus2022} and race \cite{boyd2020} have been shown to affect $D_{\text{final}}$. However, the magnitude of the differences between demographic groups is never as significant as those shown in Table~\ref{tab:d_stats}. Another possibility is that BAD's normative database might not adequately represent normal populations. That might seem unlikely given BAD's reputation for success, but if the global consensus of ophthalmology experts---including Drs. Belin and Ambrósio---agrees that ``there is no clinically adequate classification system for keratoconus'' \cite{gomes2015b}, perhaps there is no clinically adequate classification for what is \emph{not} keratoconus.

Ultimately, resolving the paradox depends on what zero and non-zero values of $D_{\text{final}}$ mean in terms of physical characteristics of the cornea. While thresholds such as $D_{\text{final}} \ge 1.6$  for "suspicious" and $D_{\text{final}} \ge 3.0$ for "abnormal" are commonly cited, there is little explanation on how to interpret the values between these thresholds or what they represent. Clinicians rely on BAD to make critical decisions, and it is essential that they have a thorough understanding of what $D_{\text{final}}$ represents and how its values should be interpreted to ensure optimal patient outcomes.

This paper will examine how BAD works to identify the underlying reason for the paradox of non-zero means. In doing so, we will identify a number of issues and limitations with BAD that are inherent to its design. We will show:

\begin{itemize}
\item
  Despite its popularity and demonstrated efficacy, BAD and its output, $D_{\text{final}}$, are not well understood.
\item
  BAD violates a number of assumptions of regression models, undermining its interpretability and its effectiveness for certain ranges of the $D$ indices.
\item
  The baseline score of the BAD model suggests the inclusion of unexplained bias, possibly affecting clinical interpretation.
\item
  Discrepancies between theoretical assumptions and empirical findings may indicate a mismatch with real-world populations, raising concerns about the adequacy of the normative database(s) used to develop BAD.~
\end{itemize}

This is a statistical analysis only. The primary sources are the cited studies and the anonymized data from a single refractive practice. Over 4,000 exam records from Pentacam version 1.29 were imported into R version 4.4.0. Only exams with a status of ``OK'' were considered. To ensure statistical independence, one randomly selected eye of each patient was chosen for this analysis. No attempt was made to distinguish between normal and abnormal eyes because the linear relationship between the $D$ indices and their source measures is independent of any normal-abnormal categorization. The final dataset contained 1,603 exams, including 777 left eyes and 826 right eyes.

\section{Overview of BAD}

BAD is a regression model that combines multiple corneal indices into a single result.

The Pentacam's help screens and documentation \cite{oculus2018} explicitly state that the model considers fives indices, $D_{\text{f}}$, $D_{\text{b}}$, $D_{\text{p}}$, $D_{\text{t}}$, and $D_{\text{y}}$.

\begin{figure}
\centering
\includegraphics[width=0.5\textwidth]{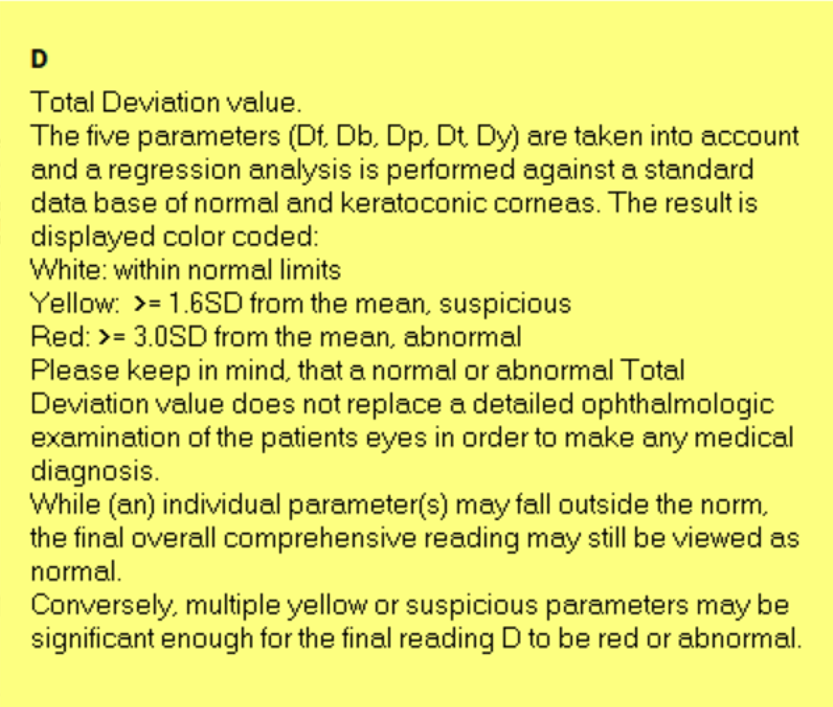}
\caption{Screenshot of the help dialog for $D_{\text{final}}$ from the ``Belin/Ambrósio Enhanced Ectasia Display''.}
\label{fig:image1}
\end{figure}

The ``Belin/Ambrósio Enhanced Ectasia Display'' likewise shows those same five indices.

\begin{figure}
\centering
\includegraphics[width=\textwidth]{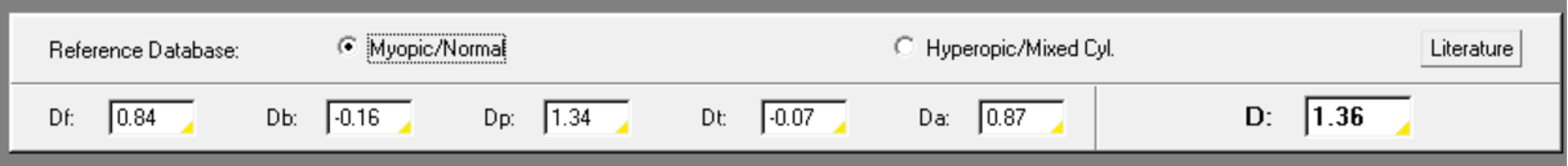}
\caption{Screenshot from the ``Belin/Ambrósio Enhanced Ectasia Display''.}
\end{figure}

However, BAD.CSV, one of the files generated by the Pentacam, contains ten indices: The five indices shown on the ``Enhanced Ectasia Display'' and five others: $D_{\text{e}}$, $D_{\text{am}}$, $D_{\text{aa}}$, $D_{\text{k}}$, and $D_{\text{r}}$.

In an Oculus-sponsored webinar, Dr. Belin shared a version of the model's regression equation that contains nine of the ten indices \cite{oculus2022}:

\begin{equation}
\small
D_{\text{final}} = w_{\text{f}}D_{\text{f}} + w_{\text{b}}D_{\text{b}} + w_{\text{p}}D_{\text{p}} + w_{\text{t}}D_{\text{t}} + w_{\text{y}}D_{\text{y}} + w_{\text{e}}D_{\text{e}} + w_{\text{am}}D_{\text{am}} + w_{\text{aa}}D_{\text{aa}} + w_{\text{k}}D_{\text{k}} + C
\label{eq:bad}
\end{equation}

Each of the indices is a z-score normalization of its source measure \cite{oculus2018}, shown in Table~\ref{tab:def}.

\begin{table}[htbp]
\caption{Definitions of the $D$ indices and their source measures.}
\label{tab:def}
\centering
\small
\resizebox{\linewidth}{!}{%
\begin{tabular}{@{}
  >{\raggedright\arraybackslash}p{0.09\linewidth}
  >{\raggedright\arraybackslash}p{0.33\linewidth}
  >{\raggedright\arraybackslash}p{0.31\linewidth}
  >{\raggedright\arraybackslash}p{0.14\linewidth}
  >{\raggedright\arraybackslash}p{0.13\linewidth}@{}}
\toprule
\textbf{Index} & \textbf{Definition} & \textbf{Source Measure} & 
\textbf{Shown on the Enhanced Ectasia Display} & 
\textbf{Included in Regression Equation} \\ \midrule
$D_{\text{f}}$ & Changes in anterior elevation between the standard and enhanced reference surfaces (BFS and eBFS) & Apex and max elevation \cite{belin2012} & Yes & Yes \\
$D_{\text{b}}$ & Changes in posterior elevation between the standard and enhanced reference surfaces (BFS and eBFS) & Apex and max elevation \cite{belin2012} & Yes & Yes \\
$D_{\text{p}}$ & Average Pachymetric Progression & rpi\_avg & Yes & Yes \\
$D_{\text{t}}$ & Corneal thickness at the thinnest point & pachy\_min & Yes & Yes \\
$D_{\text{y}}$ & Vertical displacement of the thinnest point & pachy\_min\_y & No & Yes \\
$D_{\text{e}}$ & Posterior elevation at the thinnest point & ele\_b\_bfs\_8mm\_thinnest \cite{oculus2022, ramos2012} & No & Yes \\
$D_{\text{am}}$ & Maximum Ambrósio Relational Thickness & art\_max \cite{oculus2022} & Yes (as $D_{\text{a}}$) & Yes \\
$D_{\text{aa}}$ & Average Ambrósio Relational Thickness & art\_avg \cite{oculus2022} & No & Yes \\
$D_{\text{k}}$ & K-max & k\_max\_front\_d \cite{oculus2022} & No & Yes \\
$D_{\text{r}}$ & Not explicity defined & rel\_pachy\_min & No & No \\
\end{tabular}%
}
\end{table}

Elsewhere, Drs. Belin and Ambrósio have stated that the model contains a slightly different list of nine parameters, which includes the anterior elevation at the thinnest point but excludes one of the measures of Ambrósio Relational Thickness \cite{belin2013b, belin2023}.

Given the conflicting information about parameters that comprise BAD,~we need to confirm that we have the correct $D$ indices before analyzing the model itself.

\subsection{Exploring the D Indices}

A simple way to validate that we have the correct source measures for the $D$ indices is to plot them against each other: We should be able to see a direct linear relationship between the two. This is most easily done for the indices whose source measures are exported by the Pentacam software: $D_{\text{aa}}$, $D_{\text{am}}$, $D_{\text{e}}$, $D_{\text{k}}$, $D_{\text{p}}$, $D_{\text{r}}$, $D_{\text{t}}$, and $D_{\text{y}}$.

\begin{figure}
\centering
\includegraphics[width=\textwidth]{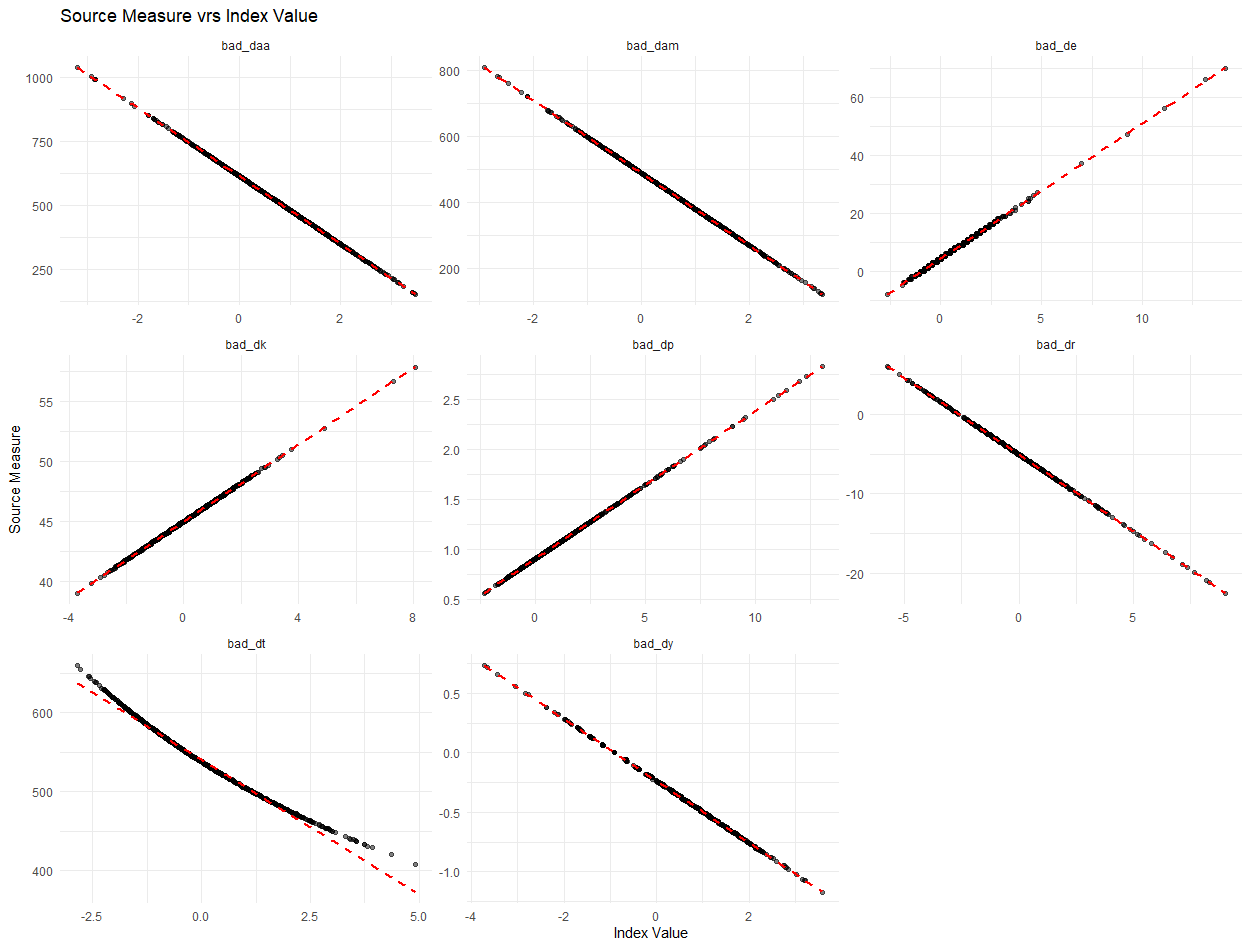} 
\caption{Scatter plots illustrating the linear relationship between the source measures and index values for $D_{\text{aa}}$, $D_{\text{am}}$, $D_{\text{e}}$, $D_{\text{k}}$, $D_{\text{p}}$, $D_{\text{r}}$, $D_{\text{t}}$, and $D_{\text{y}}$. The red dashed line illustrates the linear relationship between the source measure and index value.} \label{fig:image3}
\end{figure}

The plots in Figure~\ref{fig:image3} confirm that we have the correct sources for each of them.

$D_{\text{t}}$ is one possible exception: The line formed by the points is slightly curved, suggesting that BAD performs a subtle transformation of the source measure, but the fact that all of the points fall on the same---nearly straight---line is sufficient for our purposes.

$D_{\text{f}}$ and $D_{\text{b}}$ are more challenging because the source measurements are \emph{not} exported by the Pentacam. If we needed the original data, we would need to replicate the calculation of the enhanced Best-Fit-Sphere (eBFS) and changes in elevation between the eBFS and the standard Best-Fit-Sphere (BFS). Doing so would introduce multiple sources for potential error. Fortunately, our goal is simply to understand how BAD works, which we can accomplish with reasonable proxies for the source measures of $D_{\text{f}}$ and $D_{\text{b}}$. To that end, we made the simplifying assumption to use the absolute value of the change in radius between BFS and eBFS.

\begin{figure}
\centering
\includegraphics[width=0.75\textwidth]{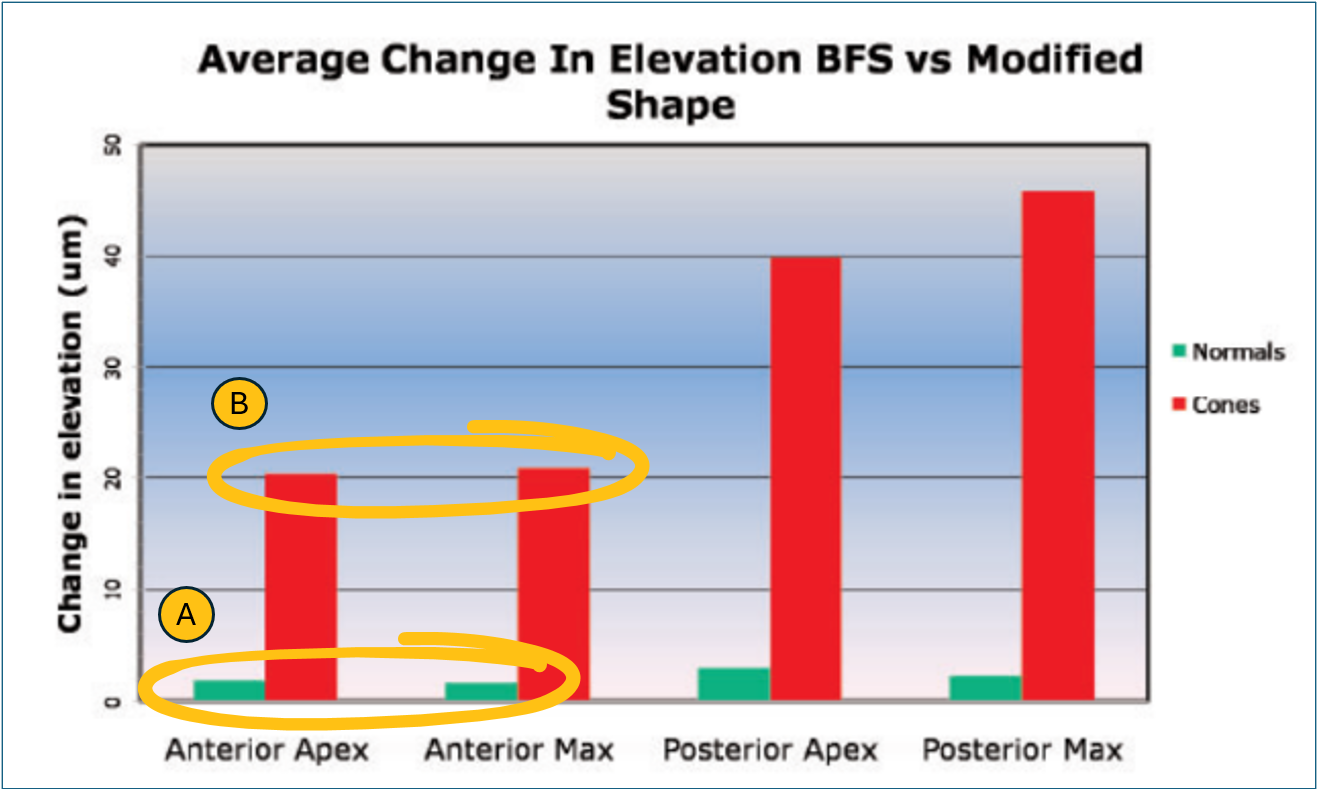}
\caption{Bar graph showing the relative change in elevation for normal eyes (green) and keratoconics (red) when comparing elevation measured with the baseline BFS and enhanced BFS. Yellow highlights added.}
\label{fig:image4}
\end{figure}

The plot in Figure~\ref{fig:image4}, adapted from published research by Drs. Belin and Ambrósio \cite{belin2012}, explains the intuition. On the anterior surface, the \emph{difference} in elevation at the apex and max locations are in the 2 -- 3 $\mu$m range for both normal eyes (green bars, highlight ``A'') and keratoconic eyes (red bars, highlight ``B''). The \emph{change} in elevation between normal and keratoconic eyes is much more pronounced: \textasciitilde20 $\mu$m at both the apex and max. The posterior surface
demonstrates the same behavior, albeit slightly more pronounced. As such, just one measure per surface should be sufficient to capture the change in elevation from BFS to eBFS. And since all measurements are taken relative to the reference shapes, the difference between the reference shapes themselves should be sufficient. One limitation, however, is that the difference measures for $D_{\text{f}}$ and $D_{\text{b}}$ are calculated at the $\mu$m grain, but the radii of the BFS and eBFS are exported at the 0.01 mm grain, resulting in a loss of precision of 10 $\mu$m.

As we did with the other source measures, we can plot our source measures against their $D$ indices to confirm the linear relationship. Although the plotted points don't form a perfectly straight lines, the correlation coefficients between our proxy measures and their $D$ indices are extremely high: 0.97 for both $D_{\text{f}}$ and $D_{\text{b}}$, confirming that their use as proxies is sufficient.

\begin{figure}
\centering
\includegraphics[width=0.75\textwidth]{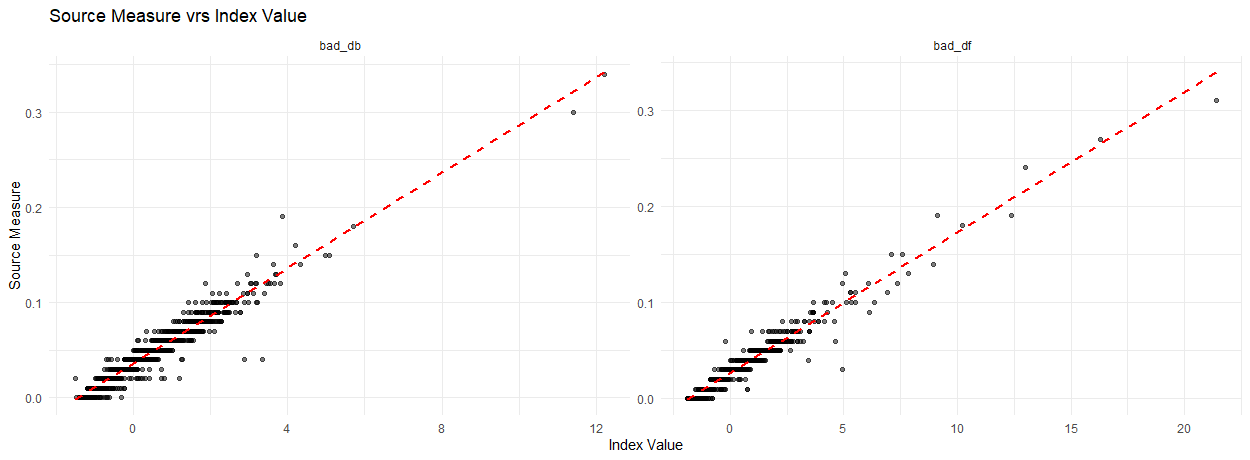}
\caption{Scatter plots illustrating the linear relationship between the source measures and index values for $D_{\text{f}}$ and $D_{\text{b}}$. The red dashed line illustrates the linear relationship between the source measure and index value.}
\label{fig:image5}
\end{figure}

\section{Z-score Normalization}

The predictors of a regression equation (e.g., the $D$ indices) are often z-score normalized to make the resulting model more interpretable. The coefficients of the fitted equation ($w_\text{i}$) reflect the change in the outcome variable ($D_{\text{final}}$) for a one-unit change in each predictor variable in the predictor's unit of measure. This makes the interpretation of each predictor dependent on its units and scale, which may not be directly comparable across predictors.

For example, the difference between ART\textsubscript{max} = 540 $\mu$m and RPI\textsubscript{avg} = 0.9 is nearly 3 orders of magnitude. If left in their original units of measure, ART\textsubscript{max} would appear to have disproportionately larger effect than RPI\textsubscript{avg} simply due to the difference in scale between predictors, not their actual impact of on $D_{\text{final}}$.

Z-score normalization standardizes all variables to the same scale, standard deviations from the mean. This allows for direct comparison of parameter effects, as each unit increase in a predictor represents a one standard deviation increase, making it easier to interpret the relative importance of each variable in the model.

Because the equation for calculating a z-score is a linear transform, it does not change the linear relationship between the predictors and the outcome:

\begin{equation}
z_{i} = \ \frac{x_{i} - \ \mu}{\sigma}
\label{eq:z-score}
\end{equation}

where:

\begin{itemize}
\item
  $x_{i}$ is the source measurement,
\item
  $\mu$ is the mean of the source measurement \emph{for normal corneas},
\item
  $\sigma$ is the standard deviation of the source measurement \emph{for normal corneas}.
\end{itemize}

\section{Thresholds for Normal}

Having both the source measures and the $D$ indices, we can identify the mean ($\mu$) and standard deviation ($\sigma$) for each of the underlying distributions.

The z-normalization formula can be rewritten in the form of a linear equation:

\begin{equation}
x_{i} = \ \mu + \ \sigma z_{i}
\label{eq:z-rewrite}
\end{equation}

This equation shows that the original value $x_{i}$ can be expressed as a linear function of the z-normalized value $z_{i}$, with:

\begin{itemize}
\item
  Intercept $\mu$ (the mean),
\item
  Slope $\sigma$ (the standard deviation).
\end{itemize}

We can apply linear regression to estimate $\mu$ and $\sigma$ by treating the original measurements $x_{i}$ as the dependent variable (response variable) and the z-normalized values $z_{i}$ as the independent variable (predictor variable). The regression model takes the form:

\begin{equation}
x_{i} = \ \beta_{0} + \ \beta_{1}z_{i} + \ \epsilon_{i}
\label{eq:z-regression}
\end{equation}

where:

\begin{itemize}
\item
  $\beta_{0}$ is the intercept, which corresponds to the mean $\mu$ in Equation~\ref{eq:z-rewrite},
\item
  $\beta_{1}$ is the slope, which corresponds to the standard deviation $\sigma$ in Equation~\ref{eq:z-rewrite},
\item
  $\epsilon_{i}$ represents the residual error (differences between the observed $x_{i}$ and the fitted values).
\end{itemize}

Table~\ref{tab:d_mu_sigma} contains the resulting mean and standard deviation of the normal population for the source measure for each of the $D$ indices.

\begin{longtable}[]{@{}
  >{\raggedright\arraybackslash}p{0.15\linewidth}
  >{\raggedright\arraybackslash}p{0.50\linewidth}
  >{\raggedleft\arraybackslash}p{0.15\linewidth}
  >{\raggedleft\arraybackslash}p{0.15\linewidth}@{}}
\caption{Mean and standard deviation for the $D$ indices based on linear regression.}
\label{tab:d_mu_sigma}
\tabularnewline
\toprule\noalign{}
\begin{minipage}[b]{\linewidth}\centering
\textbf{Index}
\end{minipage} & \begin{minipage}[b]{\linewidth}\centering
\textbf{Source Measure}
\end{minipage} & \begin{minipage}[b]{\linewidth}\centering
\textbf{Mean ($\beta_{0}$)}
\end{minipage} & \begin{minipage}[b]{\linewidth}\centering
\textbf{SD ($\beta_{1}$)}
\end{minipage} \\
\midrule\noalign{}
\endfirsthead
\toprule\noalign{}
\begin{minipage}[b]{\linewidth}\centering
\textbf{Index}
\end{minipage} & \begin{minipage}[b]{\linewidth}\centering
\textbf{Source Measure}
\end{minipage} & \begin{minipage}[b]{\linewidth}\centering
\textbf{Mean ($\beta_{0}$)}
\end{minipage} & \begin{minipage}[b]{\linewidth}\centering
\textbf{SD ($\beta_{1}$)}
\end{minipage} \\
\midrule\noalign{}
\endhead
\bottomrule\noalign{}
\endlastfoot
$D_{\text{aa}}$ & art\_avg ($\mu$m) & 614 & 133 \\
$D_{\text{am}}$ & art\_max ($\mu$m) & 488 & 109 \\
$D_{\text{b}}$ & change\_back (mm) & 0.04 & 0.03 \\
$D_{\text{e}}$ & ele\_b\_bfs\_8mm\_thinnest ($\mu$m) & 4 & 5 \\
$D_{\text{f}}$ & change\_front (mm) & 0.03 & 0.01 \\
$D_{\text{k}}$ & k\_max\_front\_d (D) & 45.0 & 1.6 \\
$D_{\text{p}}$ & rpi\_avg & 0.90 & 0.15 \\
$D_{\text{r}}$ & rel\_pachy\_min & -5.1 & 1.9 \\
$D_{\text{t}}$ & pachy\_min ($\mu$m) & 540 & 34 \\
$D_{\text{y}}$ & pachy\_min\_y (mm) & -0.24 & 0.26 \\
\end{longtable}

We can further validate the means that we've calculated by looking at exams with $D$ index values of 0. The source measures should match the $\beta_{0}$ that we've calculated. Similarly, we can validate the standard deviations by looking at exams with $D$ index values of 1. The absolute value of the difference between the source measures at index values of 1 and 0 (i.e., the ``Absolute Delta'' in Table~\ref{tab:d_empirical}) should correspond to the $\beta_{1}$ that we've calculated.

\begin{table}[htbp]
\caption{Comparison of empirical and calculated results for the mean and standard deviation of the $D$ indices. $D_{\text{f}}$ and $D_{\text{b}}$ are based on proxies for their source measures.} 
\label{tab:d_empirical}
\centering
\small
\resizebox{\linewidth}{!}{%
\begin{tabular}{@{}
  >{\raggedright\arraybackslash}p{0.10\linewidth}
  >{\raggedright\arraybackslash}p{0.30\linewidth}
  >{\centering\arraybackslash}p{0.15\linewidth}
  >{\centering\arraybackslash}p{0.15\linewidth}
  >{\centering\arraybackslash}p{0.15\linewidth}
  >{\raggedright\arraybackslash}p{0.15\linewidth}@{}}
\toprule
\textbf{Index} & \textbf{Source Measure} & \textbf{$D_{i} = 0$ (Range)} & 
\textbf{$D_{i} = 1$ (Range)} & \textbf{Absolute Delta (Range)} & 
\textbf{Direction of \newline Abnormality} \\ \midrule
$D_{\text{aa}}$ & art\_avg ($\mu$m) & (614, 615) & (481, 482) & (132 -- 134) & Decreasing \\
$D_{\text{am}}$ & art\_max ($\mu$m) & 488 & (378, 379) & (109 -- 110) & Decreasing \\
$D_{\text{b}}$ & change\_back (mm) & (0.03 -- 0.05) & (0.06) & (0.01 -- 0.03) & Increasing \\
$D_{\text{e}}$ & ele\_b\_bfs\_8mm\_thinnest & 4 & 9 & 5 & Increasing \\
$D_{\text{f}}$ & change\_front (mm) & (0.02 -- 0.03) & 0.04 & (0.01 -- 0.02) & Increasing \\
$D_{\text{k}}$ & k\_max\_front\_d (D) & 45.0 & 46.6 & 1.6 & Increasing \\
$D_{\text{p}}$ & rpi\_avg & (0.90 -- 0.91) & 1.05 & (0.14 -- 0.15) & Increasing \\
$D_{\text{r}}$ & rel\_pachy\_min & -5.1 & -7.0 & 1.9 & Decreasing \\
$D_{\text{t}}$ & pachy\_min ($\mu$m) & 538 & 505 & 33 & Decreasing \\
$D_{\text{y}}$ & pachy\_min\_y (mm) & -0.24 & -0.50 & 0.26 & Decreasing \\ \bottomrule
\end{tabular}%
}
\end{table}

In almost every case, our calculated values for the mean and standard deviation are within the empirical ranges. The one exception is $D_{\text{t}}$, which we've already identified as not having a perfectly linear relationship with its source measure. Here, the difference between the calculated and empirical parameters quantifies the difference, confirming that it is minimal.

Finally, we can use the empirical results to determine the direction of abnormality. (For example, a \emph{decreasing} min pachymetry is in the direction of abnormality.) Anytime the value of an index at $D_{i} = 1$ is less than the value of the index at $D_{i} = 0$, the direction of abnormality is decreasing. Conversely, anytime the value of an index at $D_{i} = 1$ is greater than the value of the index at $D_{i} = 0$, the direction of abnormality is increasing.

\begin{figure}
\centering
\includegraphics[width=\textwidth]{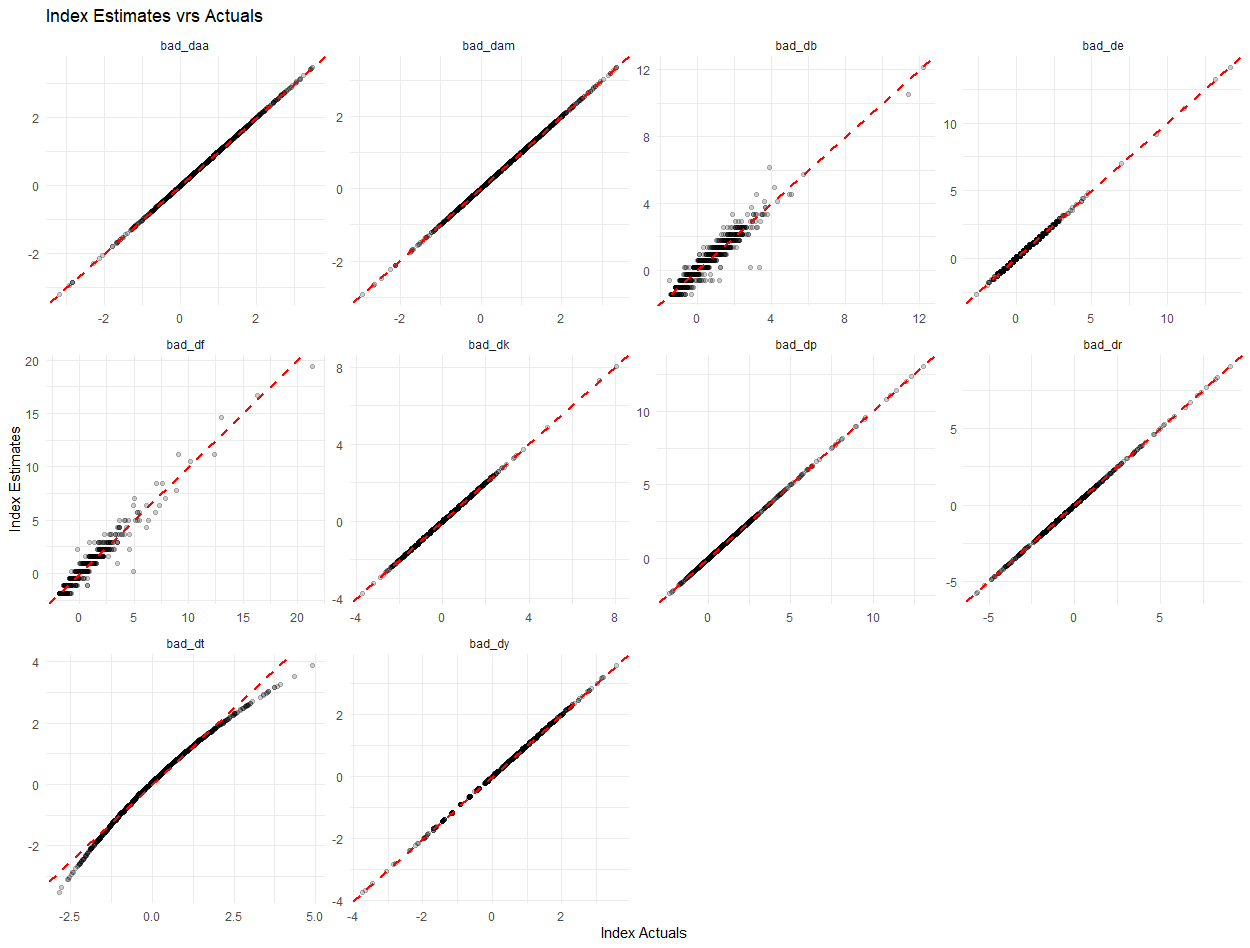}
\caption{Estimated vrs actual values of the $D$ indices. The red dashed line indicates a 1:1 relationship between estimates and actuals.}
\label{fig:image6}
\end{figure}

We can use our calculated mean and standard deviation values to normalize the source measures, effectively calculating the $D$ indices from scratch. The plots in Figure~\ref{fig:image6} compare the estimates to actuals.

The primary difference between Figure~\ref{fig:image6} and Figures~\ref{fig:image3} and \ref{fig:image5} is that all of the plot lines have a positive slope, which is a byproduct of the fact that ``the `d' values are calculated so that a value of zero represents the average of the normal population and 1 represents the value is one standard deviation towards the disease (ectasia) value'' \cite{ambrosio2013}.

We can now translate BAD's thresholds for suspicious ($\ge$ 1.6 SD) and abnormal ($\ge$ 2.6 SD) into units of the indices' source measurements, which will eventually allow us to compare the statistical parameters of our normal subgroup with the meta-analysis in the source measures' original units.

\begin{longtable}[]{@{}
  >{\raggedright\arraybackslash}p{0.05\linewidth}
  >{\raggedright\arraybackslash}p{0.30\linewidth}
  >{\raggedleft\arraybackslash}p{0.075\linewidth}  
  >{\raggedleft\arraybackslash}p{0.075\linewidth}
  >{\raggedright\arraybackslash}p{0.10\linewidth}
  >{\raggedleft\arraybackslash}p{0.10\linewidth}
  >{\raggedleft\arraybackslash}p{0.10\linewidth}@{}}
\caption{Estimated cutoff values for suspicious and abnormal exams for the $D$ indices. $D_{\text{f}}$ and $D_{\text{b}}$ are based on proxies for their source measures.}
\label{tab:d_cutoffs}
\tabularnewline
\toprule\noalign{}
\begin{minipage}[b]{\linewidth}\centering
\textbf{Index}
\end{minipage} & \begin{minipage}[b]{\linewidth}\centering
\textbf{Source Measure (units)}
\end{minipage} & \begin{minipage}[b]{\linewidth}\centering
\textbf{Mean ($\beta_{0}$)}
\end{minipage} & \begin{minipage}[b]{\linewidth}\centering
\textbf{SD ($\beta_{1}$)}\strut
\end{minipage} & \begin{minipage}[b]{\linewidth}\centering
\textbf{Direction Toward Disease}
\end{minipage} & \begin{minipage}[b]{\linewidth}\centering
\textbf{Suspicious ($\ge$ 1.6 SD)}\strut
\end{minipage} & \begin{minipage}[b]{\linewidth}\centering
\textbf{Abnormal ($\ge$ 2.6 SD)}\strut
\end{minipage} \\
\midrule\noalign{}
\endfirsthead
\toprule\noalign{}
\begin{minipage}[b]{\linewidth}\centering
\textbf{Index}
\end{minipage} & \begin{minipage}[b]{\linewidth}\centering
\textbf{Source Measure (units)}
\end{minipage} & \begin{minipage}[b]{\linewidth}\centering
\textbf{Mean ($\beta_{0}$)}
\end{minipage} & \begin{minipage}[b]{\linewidth}\centering
\textbf{SD ($\beta_{1}$)}\strut
\end{minipage} & \begin{minipage}[b]{\linewidth}\centering
\textbf{Direction Toward Disease}
\end{minipage} & \begin{minipage}[b]{\linewidth}\centering
\textbf{Suspicious ($\ge$ 1.6 SD)}\strut
\end{minipage} & \begin{minipage}[b]{\linewidth}\centering
\textbf{Abnormal ($\ge$ 2.6 SD)}\strut
\end{minipage} \\
\midrule\noalign{}
\endhead
\bottomrule\noalign{}
\endlastfoot
$D_{\text{aa}}$ & art\_avg ($\mu$m) & 614 & 133 & Decreasing & 401 & 269 \\
$D_{\text{am}}$ & art\_max ($\mu$m) & 488 & 109 & Decreasing & 313 & 203 \\
$D_{\text{b}}$ & change\_back (mm) & 0.04 & 0.03 & Increasing & 0.08 & 0.10 \\
$D_{\text{e}}$ & ele\_b\_bfs\_8mm\_thinnest ($\mu$m) & 4 & 5 & Increasing & 12 &
16 \\
$D_{\text{f}}$ & change\_front (mm) & 0.03 & 0.01 & Increasing & 0.05 & 0.06 \\
$D_{\text{k}}$ & k\_max\_front\_d (D) & 45.0 & 1.6 & Increasing & 47.6 &
49.2 \\
$D_{\text{p}}$ & rpi\_avg & 0.90 & 0.15 & Increasing & 1.14 & 1.29 \\
$D_{\text{r}}$ & rel\_pachy\_min & -5.1 & 1.9 & Decreasing & -8.2 & -10.1 \\
$D_{\text{t}}$ & pachy\_min ($\mu$m) & 540 & 34 & Decreasing & 486 & 452 \\
$D_{\text{y}}$ & pachy\_min\_y (mm) & -0.24 & 0.26 & Decreasing & -0.65 &
-0.91 \\
\end{longtable}

\section{Comparing Distributions}

Knowing the parameters that define the normal-group distributions of the $D$ indices, we can compare them to the probability density functions (PDFs) of our sample group. To simplify the comparison, we will plot the PDFs of our sample group's $D$ values relative to the standard normal distribution, the equivalent to BAD's z-score normalized reference datasets.

\begin{figure}
\centering
\includegraphics[width=\textwidth]{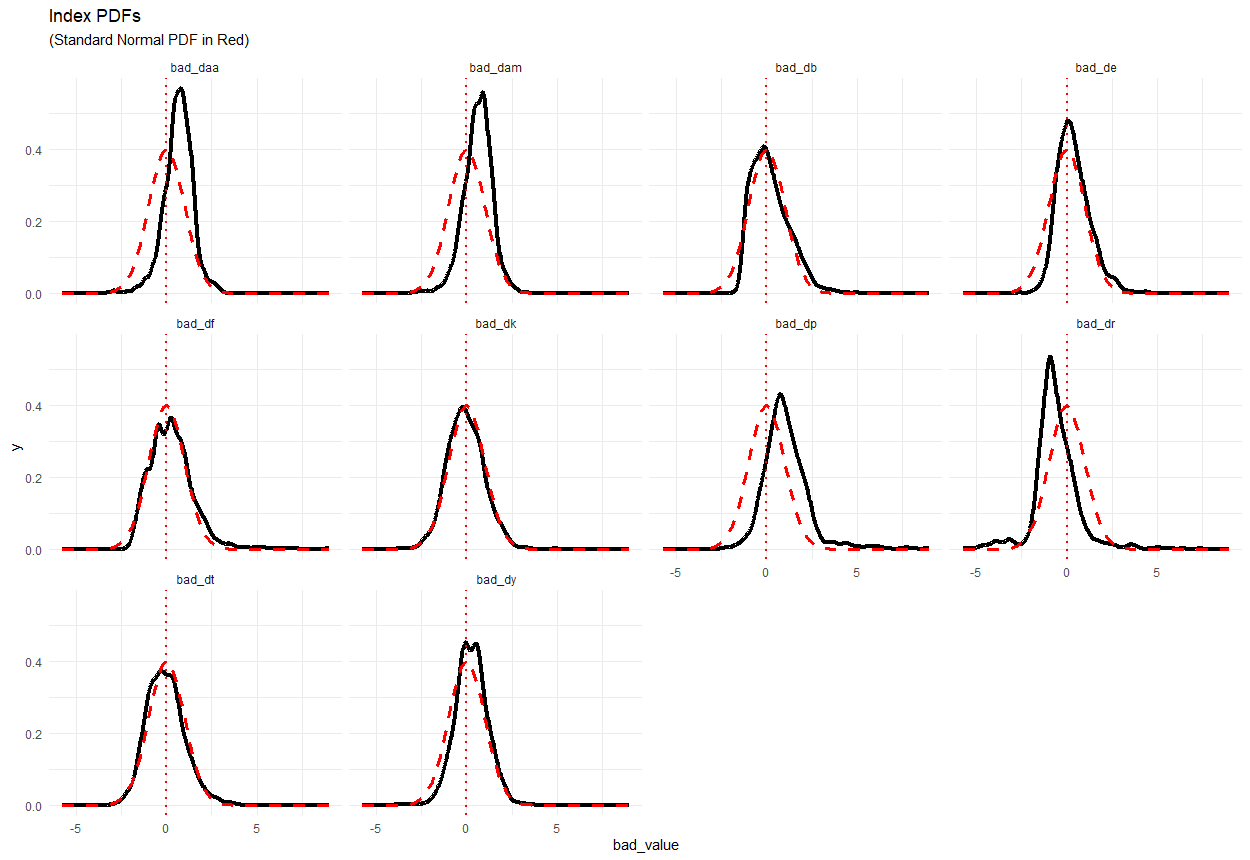}
\caption{Empirical density curves for the $D$ indices. The red dashed line is the standard normal distribution, and the red dotted line is the mode of the standard normal distribution.}
\label{fig:image7}
\end{figure}

Most of the sample's PDFs align with the standard normal distribution: $D_{\text{b}}$, $D_{\text{e}}$, $D_{\text{f}}$, $D_{\text{k}}$, $D_{\text{t}}$, and $D_{\text{y}}$ are all reasonably close. However, $D_{\text{aa}}$, $D_{\text{am}}$, and $D_{\text{p}}$ are shifted to the right, and $D_{\text{r}}$ is shifted to the left.

One useful characteristic of the normal distribution is that its mean, median, and mode are identical. When a distribution is perfectly normal, all three are located at the peak of the bell curve. When a distribution is skewed, however, the mean and median will shift in the direction of the outliers. But not the mode: The mode is generally less affected by skew because it represents the most frequently occurring value in the dataset, which should remain stable (depending on the nature of the skew). As such, we can quantify the shift in these distributions by looking at their modes.

\begin{longtable}[]{@{}
  >{\raggedright\arraybackslash}p{0.10\linewidth}
  >{\raggedleft\arraybackslash}p{0.10\linewidth}|
  >{\raggedright\arraybackslash}p{0.40\linewidth}
  >{\raggedleft\arraybackslash}p{0.10\linewidth}@{}}
\caption{Empirical modes of the $D$ indices and their source
measures.}
\label{tab:d_modes}
\tabularnewline
\toprule\noalign{}
\begin{minipage}[b]{\linewidth}\raggedright
\textbf{Index}
\end{minipage} & \begin{minipage}[b]{\linewidth}\raggedleft
\textbf{Mode}
\end{minipage} & \begin{minipage}[b]{\linewidth}\raggedright
\textbf{Source Measure}
\end{minipage} & \begin{minipage}[b]{\linewidth}\raggedleft
\textbf{Mode}
\end{minipage} \\
\midrule\noalign{}
\endfirsthead
\toprule\noalign{}
\begin{minipage}[b]{\linewidth}\raggedright
\textbf{Index}
\end{minipage} & \begin{minipage}[b]{\linewidth}\raggedleft
\textbf{Mode}
\end{minipage} & \begin{minipage}[b]{\linewidth}\raggedright
\textbf{Source Measure}
\end{minipage} & \begin{minipage}[b]{\linewidth}\raggedleft
\textbf{Mode}
\end{minipage} \\
\midrule\noalign{}
\endhead
\bottomrule\noalign{}
\endlastfoot
$D_{\text{aa}}$ & 0.78 & art\_avg ($\mu$m) & 511 \\
$D_{\text{am}}$ & 0.89 & art\_max ($\mu$m) & 390 \\
$D_{\text{b}}$ & -0.16 & change\_back (mm) & 0.04 \\
$D_{\text{e}}$ & 0.07 & ele\_b\_bfs\_8mm\_thinnest ($\mu$m) & 5 \\
$D_{\text{f}}$ & 0.26 & change\_front (mm) & 0.03 \\
$D_{\text{k}}$ & -0.21 & k\_max\_front\_d (D) & 44.7 \\
$D_{\text{p}}$ & 0.76 & rpi\_avg & 1.02 \\
$D_{\text{r}}$ & -0.93 & rel\_pachy\_min & -3.3 \\
$D_{\text{t}}$ & -0.27 & pachy\_min ($\mu$m) & 529 \\
$D_{\text{y}}$ & -0.05 & pachy\_min\_y (mm) & -0.22 \\
\end{longtable}

The modes of $D_{\text{aa}}$, $D_{\text{am}}$, and $D_{\text{p}}$ are 0.78, 0.89, and 0.76 standard deviations from the mean, respectively. The mode of $D_{\text{r}}$ is -0.93 standard deviations from the mean.

\begin{figure}
\centering
\includegraphics[width=\textwidth]{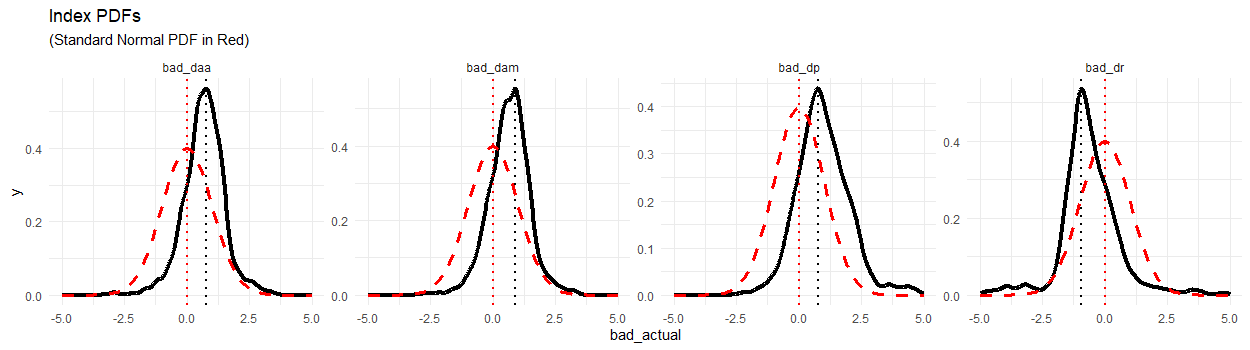}
\caption{Empirical density curves of $D_{\text{aa}}$, $D_{\text{am}}$, $D_{\text{p}}$, and $D_{\text{r}}$. The black dotted line is the mode of the density curve, the red dashed line is the standard normal distribution, and the red dotted line is the mode of the standard normal distribution.}
\label{fig:image8}
\end{figure}

Figure~\ref{fig:image8} shows the PDFs with black dashed lines drawn at their modes.

For this analysis, we did not distinguish between normal and abnormal corneas, so our sample includes corneas that BAD would categorize as normal ($<$ 1.6 SD), suspicious ($\ge$ 1.6 SD), and abnormal ($\ge$ 2.6 SD). Arguably, however, any random sample of normal corneas \emph{should} include all three categories. BAD models each of the indices as a standard normal distribution, so 94.5\% of exams should fall within the normal range, 5.0\% within suspicious, and 0.4\% within abnormal.

Table~\ref{tab:d_distrib} shows the percentage of exams that fall within each of the three categories for each of the $D$ indices. In most cases the percentages are within a few points of expectation. The differences are only enough to affect the widths of the bell-shaped curves, not the locations of their peaks.

\begin{table}[htbp]
\caption{Distribution of categories for the $D$ indices.}
\label{tab:d_distrib}
\centering
\small
\resizebox{\linewidth}{!}{%
\begin{tabular}{@{}
  >{\raggedright\arraybackslash}p{0.114\linewidth}
  >{\raggedleft\arraybackslash}p{0.0805\linewidth}
  >{\raggedleft\arraybackslash}p{0.0913\linewidth}
  >{\raggedleft\arraybackslash}p{0.0955\linewidth}
  >{\raggedleft\arraybackslash}p{0.0773\linewidth}
  >{\raggedleft\arraybackslash}p{0.0773\linewidth}
  >{\raggedleft\arraybackslash}p{0.0773\linewidth}
  >{\raggedleft\arraybackslash}p{0.0774\linewidth}
  >{\raggedleft\arraybackslash}p{0.0773\linewidth}
  >{\raggedleft\arraybackslash}p{0.0773\linewidth}
  >{\raggedleft\arraybackslash}p{0.0773\linewidth}
  >{\raggedleft\arraybackslash}p{0.0774\linewidth}@{}}
\toprule
\textbf{Category} & \textbf{Target} & \textbf{$D_{\text{aa}}$} & 
\textbf{$D_{\text{am}}$} & \textbf{$D_{\text{b}}$} & \textbf{$D_{\text{e}}$} & 
\textbf{$D_{\text{f}}$} & \textbf{$D_{\text{k}}$} & \textbf{$D_{\text{p}}$} & 
\textbf{$D_{\text{r}}$} & \textbf{$D_{\text{t}}$} & \textbf{$D_{\text{y}}$} \\ \midrule
Normal & 94.5\% & 90.9 & 91.5 & 89.0 & 90.1 & 87.1 & 93.8 & 73.4 & 95.6 & 92.9 & 93.4 \\
Suspicious & 5.0\% & 7.3 & 7.2 & 8.5 & 7.1 & 8.5 & 5.6 & 18.8 & 1.9 & 5.2 & 5.9 \\
Abnormal & 0.4\% & 1.8 & 1.3 & 2.5 & 2.8 & 4.4 & 0.7 & 7.8 & 5.2 & 1.9 & 0.6 \\
Total & 100\% & 100.0 & 100.0 & 100.0 & 100.0 & 100.0 & 100.0 & 100.0 & 100.0 & 100.0 & 100.0 \\ \bottomrule
\end{tabular}%
}
\end{table}

Revisiting the meta-analysis confirms that most (but not all) of the published studies reported higher means (and/or medians) for $D_{\text{aa}}$, $D_{\text{am}}$, and $D_{\text{p}}$.

\begin{table}[htbp]
\caption{Statistical parameters of $D_{\text{aa}}$ in published studies. Measurements reported in the source unit of measure (shaded grey) have been converted to SD. *The order of the converted range has been reversed to maintain consistency with other ranges in the table.}
\label{tab:d_aa}
\centering
\small
\resizebox{\linewidth}{!}{%
\begin{tabular}{@{}
  >{\centering\arraybackslash}p{0.20\linewidth}
  >{\centering\arraybackslash}p{0.10\linewidth}
  >{\centering\arraybackslash}p{0.20\linewidth}
  >{\centering\arraybackslash}p{0.20\linewidth}
  >{\centering\arraybackslash}p{0.10\linewidth}
  >{\centering\arraybackslash}p{0.10\linewidth}@{}}
\toprule
\textbf{Mean ± SD} & 
\textbf{Median} & 
\textbf{Range} & 
\textbf{Group Size (Normal/ Control)} & 
\textbf{Notes} & 
\textbf{Study [Source]} \\ 
\midrule
0.44 ± 0.71 & & & 100 & & \cite{hashemi2016} \\
\cellcolor[gray]{0.9} 555 ± 94 & & & & & \\
\midrule
-0.32 ± 0.78 & -0.24 & (-2.90, 1.96) & 200 & & \cite{ramos2012} \\
\midrule
0.50 ± 0.88 & 0.46 & (-0.09, 0.93)* & 196 & IQR & \cite{steinberg2015} \\
\cellcolor[gray]{0.9} 548.2 ± 116.8 & \cellcolor[gray]{0.9} 553.0 & \cellcolor[gray]{0.9} (490, 626) & & & \\
\bottomrule
\end{tabular}%
}
\end{table}

\begin{table}[htbp]
\caption{: Statistical parameters of $D_{\text{am}}$ in published studies. Measurements reported in the source unit of measure (shaded grey) have been converted to SD. *The order of the converted range has been reversed to maintain consistency with other ranges in the table.}
\label{tab:d_am}
\centering
\small
\resizebox{\linewidth}{!}{%
\begin{tabular}{@{}
  >{\centering\arraybackslash}p{0.20\linewidth}
  >{\centering\arraybackslash}p{0.10\linewidth}
  >{\centering\arraybackslash}p{0.20\linewidth}
  >{\centering\arraybackslash}p{0.20\linewidth}
  >{\centering\arraybackslash}p{0.10\linewidth}
  >{\centering\arraybackslash}p{0.10\linewidth}@{}}
\toprule
\textbf{Mean ± SD} & 
\textbf{Median} & 
\textbf{Range} & 
\textbf{Group Size (Normal/ Control)} & 
\textbf{Notes} & 
\textbf{Study [Source]} \\ 
\midrule
& 0.23 & (-0.08, 0.64) & 137 & IQR & \cite{song2023} \\
\midrule
& 0.64 & (0.20, 0.97) & 200 & IQR & \cite{ding2024} \\
\midrule
0.36 ± 0.75 & & & 100 & & \cite{hashemi2016} \\
\cellcolor[gray]{0.9} 449 ± 82 & & & & & \\
\midrule
-0.35 ± 0.83 & -0.22 & (-3.84, 2.46) & 200 & & \cite{ramos2012} \\
\midrule
0.14 ± 0.77 & & (-3.70, 2.15)* & 682 & & \cite{villavicencio2014} \\
\cellcolor[gray]{0.9} 473 ± 84.3 & & \cellcolor[gray]{0.9} (253, 893) & & & \\
\midrule
0.21 ± 0.65 & & & 70 & & \cite{toprak2023} \\
\midrule
0.46 ± 0.91 & 0.41 & (-0.13, 0.85)* & 196 & IQR & \cite{steinberg2015} \\
\cellcolor[gray]{0.9} 437.6±100.1 & \cellcolor[gray]{0.9} 443.5 & \cellcolor[gray]{0.9} (395, 502) & & & \\
\midrule
& 0.49 & (-0.01, 1.00)* & 42 & & \cite{shetty2017} \\
& \cellcolor[gray]{0.9} 434.5 & \cellcolor[gray]{0.9} (379, 489) & & & \\
\bottomrule
\end{tabular}%
}
\end{table}

\begin{table}[htbp]
\caption{: Statistical parameters of $D_{\text{p}}$ in published studies. Measurements reported in the source unit of measure (shaded grey) have been converted to SD.}
\label{tab:d_p}
\centering
\small
\resizebox{\linewidth}{!}{%
\begin{tabular}{@{}
  >{\centering\arraybackslash}p{0.20\linewidth}
  >{\centering\arraybackslash}p{0.10\linewidth}
  >{\centering\arraybackslash}p{0.20\linewidth}
  >{\centering\arraybackslash}p{0.20\linewidth}
  >{\centering\arraybackslash}p{0.10\linewidth}
  >{\centering\arraybackslash}p{0.10\linewidth}@{}}
\toprule
\textbf{Mean ± SD} & 
\textbf{Median} & 
\textbf{Range} & 
\textbf{Group Size (Normal/ Control)} & 
\textbf{Notes} & 
\textbf{Study [Source]} \\ 
\midrule
& 0.62 & (0.03, 1.30) & 137 & IQR & \cite{song2023} \\
\midrule
& 0.86 & (0.40, 1.38) & 200 & IQR & \cite{ding2024} \\
\midrule
0.58 ± 0.95 & & & 100 & & \cite{hashemi2016} \\
\cellcolor[gray]{0.9} 0.99 ± 0.14 & & & & & \\
\midrule
-0.35 ± 0.72 & -0.39 & (-2.08, 2.56) & 200 & & \cite{ramos2012} \\
\midrule
0.10 ± 0.88 & & (-2.54, 3.08) & 682 & & \cite{villavicencio2014} \\
\cellcolor[gray]{0.9} 0.92 ± 0.13 & & \cellcolor[gray]{0.9} (0.53, 1.36) & & & \\
\midrule
0.35 ± 0.65 & & & 70 & & \cite{toprak2023} \\
\midrule
0.64 ± 2.03 & 0.64 & (-0.03, 1.32) & 196 & IQR & \cite{steinberg2015} \\
\cellcolor[gray]{0.9} 1.0 ± 0.3 & \cellcolor[gray]{0.9} 1.0 & \cellcolor[gray]{0.9} (0.9, 1.1) & & & \\
\midrule
& 0.51 & (-0.17, 1.12) & 42 & & \cite{shetty2017} \\
& \cellcolor[gray]{0.9} 0.98 & \cellcolor[gray]{0.9} (0.88, 1.07) & & & \\
\bottomrule
\end{tabular}%
}
\end{table}

We cannot compare our $D_{\text{r}}$ to published studies, however, because that parameter is generally not included in published studies.

\subsection{Multicollinearity}

Of course, it's not surprising that $D_{\text{aa}}$, $D_{\text{am}}$, and $D_{\text{p}}$ all behave similarly. $D_{\text{aa}}$ and $D_{\text{am}}$ are the average and maximum of the same index, Ambrósio Relational Thickness (ART), which is inversely related to the Pachymetric Progression Index (PPI) \cite{ambrosio2011, ambrosio2011b}:

\begin{equation}
ART = \frac{T_{\text{p}}}{PPI}
\label{eq:art}
\end{equation}

where:

\begin{itemize}
\item
  $ART$ is the Ambrósio Relational Thickness,
\item
  $T_{\text{P}}$ is the thinnest pachymetry,
\item
  $PPI$ is the Pachymetric Progression Index.
\end{itemize}

The inherent relationship between $D_{\text{aa}}$, $D_{\text{am}}$, and $D_{\text{p}}$ is a source of potential concern. Regression models assume that there is no perfect multicollinearity between variables, which would make it impossible to estimate unique coefficients for each predictor. But even high (but less than perfect) multicollinearity can affect a model: It can inflate standard errors of coefficients, making it harder to determine the significance of individual predictors.

\begin{figure}
\centering
\includegraphics[width=\textwidth]{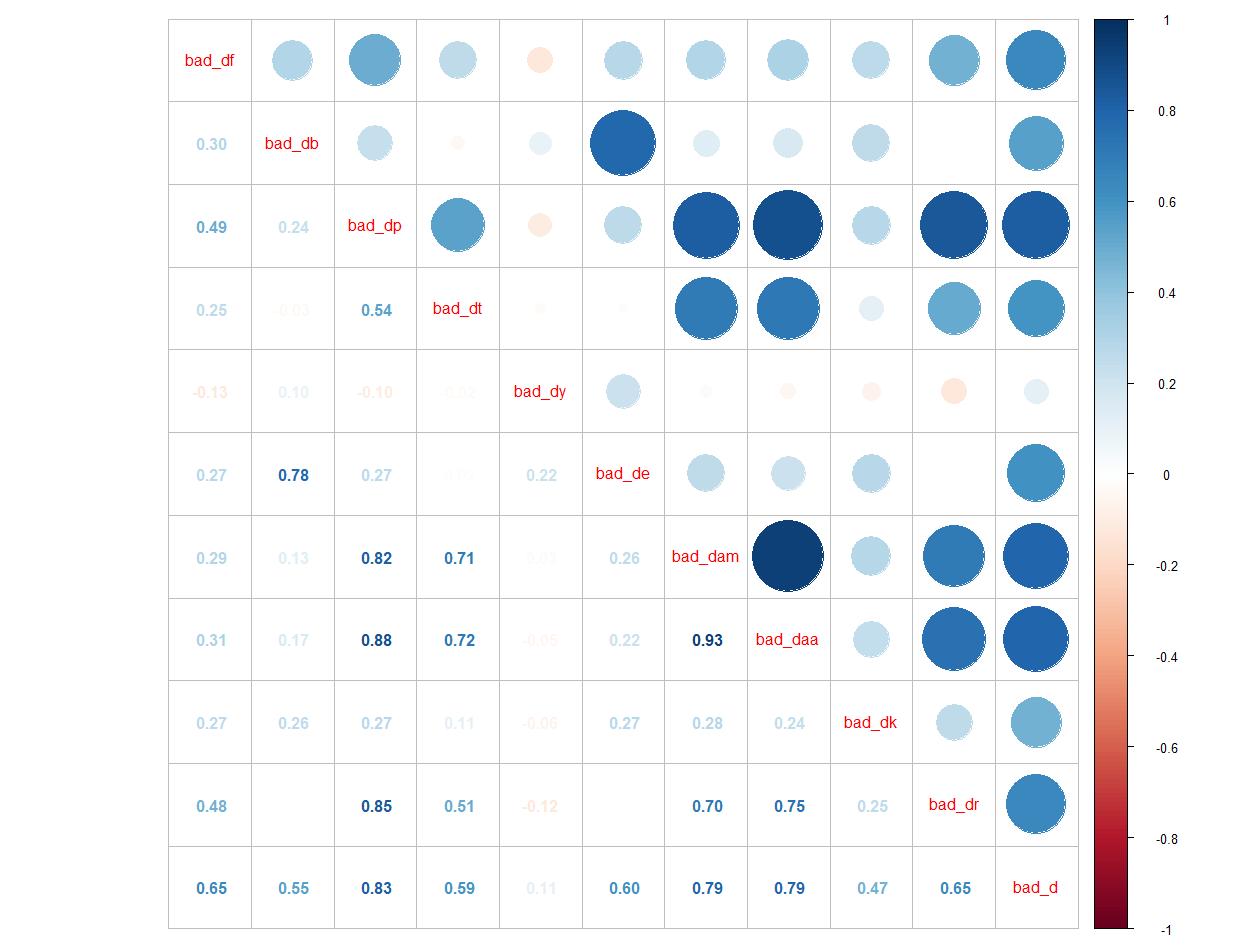}
\caption{Correlation matrix of the $D$ indices and D\textsubscript{final}.}
\label{fig:image9}
\end{figure}

Figure~\ref{fig:image9} shows the correlation between each of the $D$ indices and
$D_{\text{final}}$.

As expected, $D_{\text{aa}}$, $D_{\text{am}}$, and $D_{\text{p}}$ are highly correlated with each other, with correlation coefficients of 0.93, 0.88, and 0.82.

Based on Equation~\ref{eq:art}, $D_{\text{aa}}$ and $D_{\text{am}}$ should also be highly correlated with $D_{\text{t}}$. And, indeed, the correlation coefficients are 0.71 and 0.72.

\begin{figure}
\centering
\includegraphics[width=\textwidth]{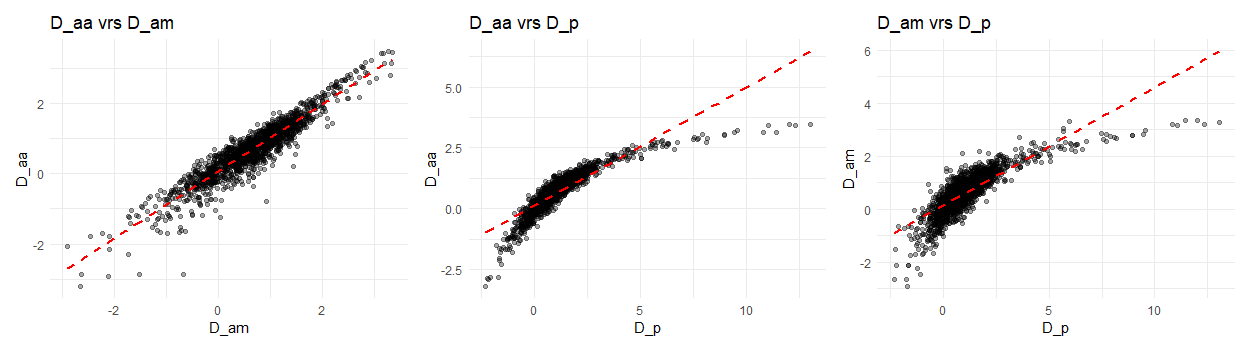}
\caption{Scatter plots illustrating the correlation between $D_{\text{aa}}$, $D_{\text{am}}$ , and $D_{\text{p}}$.}
\label{fig:image10}
\end{figure}

To illustrate the strength of these correlations, Figure~\ref{fig:image10} shows $D_{\text{aa}}$, $D_{\text{am}}$ , and $D_{\text{p}}$ plotted against each other, but it also reveals another potential issue: $D_{\text{aa}}$ and $D_{\text{am}}$ have a curvilinear relationship with $D_{\text{p}}$. The red dashed lines illustrate that, when there's a non-linear relationship exists between variables, a linear model cannot fully capture that relationship.

Moreover, the curvilinear relationship between $D_{\text{aa}}$ and $D_{\text{am}}$ and $D_{\text{p}}$ means that---at a minimum---either $D_{\text{aa}}$ and $D_{\text{am}}$ or $D_{\text{p}}$ have a curvilinear relationship with $D_{\text{final}}$.

Plotting all of the $D$ indices against $D_{\text{final}}$ confirms that $D_{\text{aa}}$, $D_{\text{am}}$, and $D_{\text{p}}$---and potentially $D_{\text{f}}$---have curvilinear relationships with $D_{\text{final}}$. Moreover, the increasing variance between $D_{\text{e}}$ and $D_{\text{final}}$ at higher values of either variable violates another fundamental assumption of regression models.

In either case---if the relationship between a predictor and the output variable is non-linear or heteroscedastic---the model's predictions risk being less accurate for certain ranges of the predictor, reducing the model\textquotesingle s overall validity. This might explain why BAD has been observed to be less accurate for specific ranges of certain characteristics, such as the white-to-white corneal diameter \cite{ding2020, lin2022}, although further analysis is needed to consider that possibility.

\begin{figure}
\centering
\includegraphics[width=\textwidth]{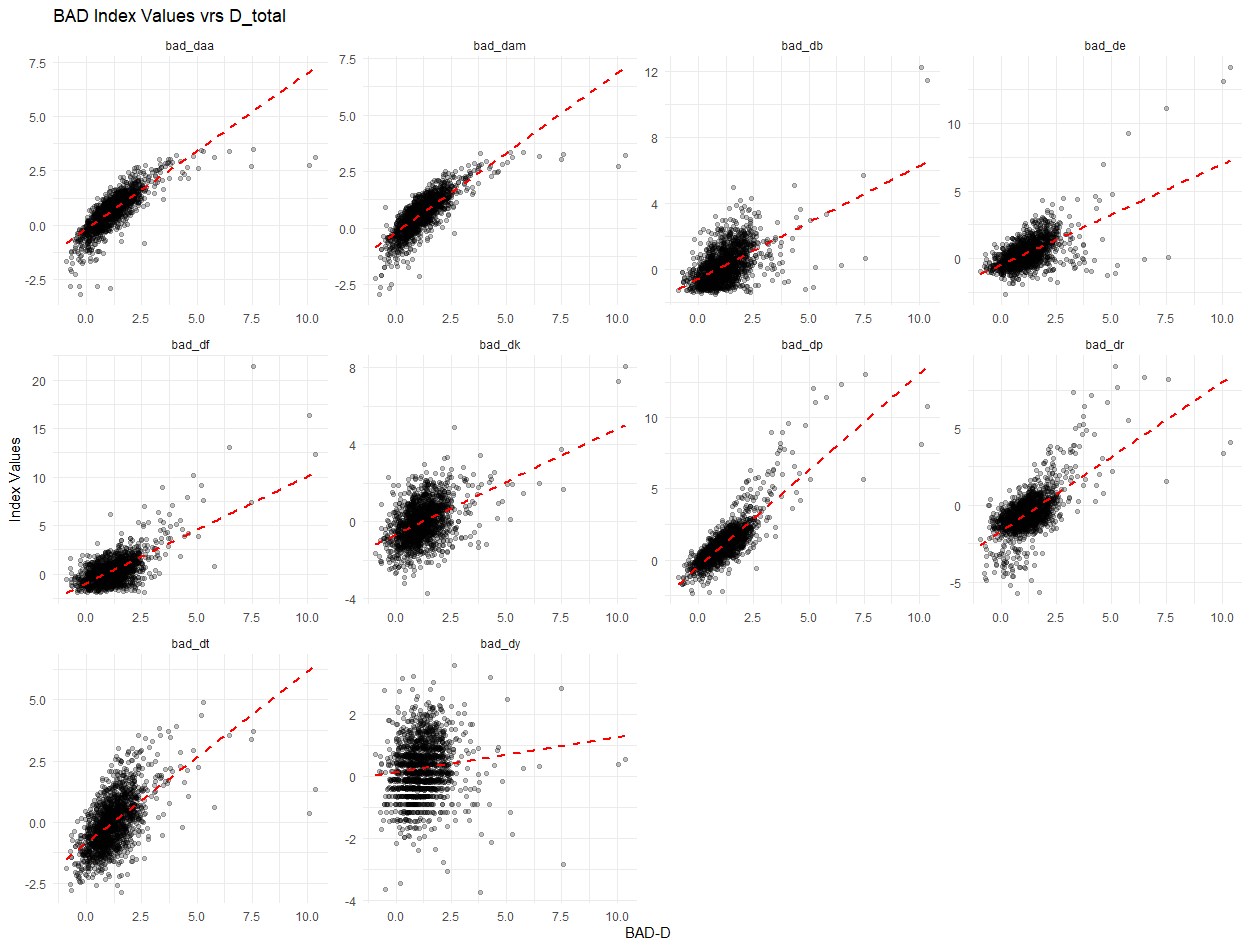} 
\caption{Scatter plots illustrating the correlation between $D$ indices and $D_{\text{final}}$. The red dashed line represents the linear relationship between the index values and $D_{\text{final}}$.}
\label{fig:image11}
\end{figure}

\section{Linear Regression}

Equation~\ref{eq:bad} says that $D_{\text{final}}$ is the weighted sum of nine $D$ indices, each of which is measure in standard deviations from the mean. Despite the fact each of the $D$ indices has its own mean and standard deviation, their sum ($D_{\text{final}}$) is likewise measured in standard deviations.

An important caveat, however, is that $D_{\text{final}}$ is an abstract measure in terms of standard deviations. The concept is similar to adding percentages. If you add percentages that are based on different denominators, the sum is still a percentage, but it is an abstract measure that has no direct connection to a specific physical quantity.

This has at least one important consequence: $D_{\text{final}}$ cannot be interpreted directly as a z-score. Conceptually, $D_{\text{final}}$ represents the cumulative distance from the mean across \emph{multiple} dimensions. The result will follow a normal distribution, but its standard deviation won't be one (1) like a typical z-score. The standard deviation of $D_{\text{final}}$ depends on both the weights of the regression equation $w_{i}$ and the correlations between the $D$ indices ($\rho_{ij}$):

\begin{equation}
SD_{\text{final}} = \sqrt{\sum_{i=1}^{n} w_{i}^2 + 2 \sum_{i<j} w_{i}w_{j}\rho_{ij}}
\label{eq:sd_final}
\end{equation}

This might explain why $D_{\text{final}}$ has a different threshold for abnormal ($\ge$ 3.0 SD) than the $D$ indices ($\ge$ 2.6 SD). Other possibilities exist, however, which we will explore later.

As with percentages, though, zero remains zero. If each of the $D$ indices has a mean of zero, the mean of their sum is also zero. Ignoring the constant ($C$) in BAD's regression equation for the moment, if every characteristic of an exam were statistically normal, all of the $D$ indices would be zero, and their sum---$D_{\text{final}}$---would be zero, regardless of the weights ($w_\text{i}$) in the equation. For the mean of $D_{\text{final}}$ to be non-zero, one or more of the $D$ indices would have to be non-zero.

We cannot ignore $C$, however.

To estimate $C$, we can fit a linear model based on Equation~\ref{eq:bad}. Table~\ref{tab:d_coef} contains the resulting $C$ (or intercept) and weights (or coefficients) for each of the $D$ indices.

\begin{longtable}[]{@{}
  >{\raggedright\arraybackslash}p{(\linewidth - 18\tabcolsep) * \real{0.1000}}
  >{\raggedleft\arraybackslash}p{(\linewidth - 18\tabcolsep) * \real{0.1000}}
  >{\raggedleft\arraybackslash}p{(\linewidth - 18\tabcolsep) * \real{0.1000}}
  >{\raggedleft\arraybackslash}p{(\linewidth - 18\tabcolsep) * \real{0.1000}}
  >{\raggedleft\arraybackslash}p{(\linewidth - 18\tabcolsep) * \real{0.1000}}
  >{\raggedleft\arraybackslash}p{(\linewidth - 18\tabcolsep) * \real{0.1000}}
  >{\raggedleft\arraybackslash}p{(\linewidth - 18\tabcolsep) * \real{0.1000}}
  >{\raggedleft\arraybackslash}p{(\linewidth - 18\tabcolsep) * \real{0.1000}}
  >{\raggedleft\arraybackslash}p{(\linewidth - 18\tabcolsep) * \real{0.1000}}
  >{\raggedleft\arraybackslash}p{(\linewidth - 18\tabcolsep) * \real{0.1000}}@{}}
\caption{Coefficients of the BAD regression equation.}
\label{tab:d_coef}
\tabularnewline
\toprule\noalign{}
\begin{minipage}[b]{\linewidth}\raggedright
\emph{\textbf{C}}
\end{minipage} & \begin{minipage}[b]{\linewidth}\raggedleft
\emph{\textbf{w\textsubscript{aa}}}
\end{minipage} & \begin{minipage}[b]{\linewidth}\raggedleft
\emph{\textbf{w\textsubscript{am}}}
\end{minipage} & \begin{minipage}[b]{\linewidth}\raggedleft
\emph{\textbf{w\textsubscript{b}}}
\end{minipage} & \begin{minipage}[b]{\linewidth}\raggedleft
\emph{\textbf{w\textsubscript{e}}}
\end{minipage} & \begin{minipage}[b]{\linewidth}\raggedleft
\emph{\textbf{w\textsubscript{f}}}
\end{minipage} & \begin{minipage}[b]{\linewidth}\raggedleft
\emph{\textbf{w\textsubscript{k}}}
\end{minipage} & \begin{minipage}[b]{\linewidth}\raggedleft
\emph{\textbf{w\textsubscript{p}}}
\end{minipage} & \begin{minipage}[b]{\linewidth}\raggedleft
\emph{\textbf{w\textsubscript{t}}}
\end{minipage} & \begin{minipage}[b]{\linewidth}\raggedleft
\emph{\textbf{w\textsubscript{y}}}
\end{minipage} \\
\midrule\noalign{}
\endfirsthead
\toprule\noalign{}
\begin{minipage}[b]{\linewidth}\raggedright
\emph{\textbf{C}}
\end{minipage} & \begin{minipage}[b]{\linewidth}\raggedleft
\emph{\textbf{w\textsubscript{aa}}}
\end{minipage} & \begin{minipage}[b]{\linewidth}\raggedleft
\emph{\textbf{w\textsubscript{am}}}
\end{minipage} & \begin{minipage}[b]{\linewidth}\raggedleft
\emph{\textbf{w\textsubscript{b}}}
\end{minipage} & \begin{minipage}[b]{\linewidth}\raggedleft
\emph{\textbf{w\textsubscript{e}}}
\end{minipage} & \begin{minipage}[b]{\linewidth}\raggedleft
\emph{\textbf{w\textsubscript{f}}}
\end{minipage} & \begin{minipage}[b]{\linewidth}\raggedleft
\emph{\textbf{w\textsubscript{k}}}
\end{minipage} & \begin{minipage}[b]{\linewidth}\raggedleft
\emph{\textbf{w\textsubscript{p}}}
\end{minipage} & \begin{minipage}[b]{\linewidth}\raggedleft
\emph{\textbf{w\textsubscript{t}}}
\end{minipage} & \begin{minipage}[b]{\linewidth}\raggedleft
\emph{\textbf{w\textsubscript{y}}}
\end{minipage} \\
\midrule\noalign{}
\endhead
\bottomrule\noalign{}
\endlastfoot
0.640 & 0.133 & 0.132 & 0.140 & 0.158 & 0.154 & 0.132 & 0.166 & 0.168 &
0.132 \\
\end{longtable}

The adjusted $R^2$ of the model is one (1), meaning that it fits the data perfectly. As such, we will accept that Equation~\ref{eq:bad} is the correct version of the BAD regression equation, and we will ignore $D_{\text{r}}$ for the remainder of this analysis.

\begin{figure}
\centering
\includegraphics[width=\textwidth]{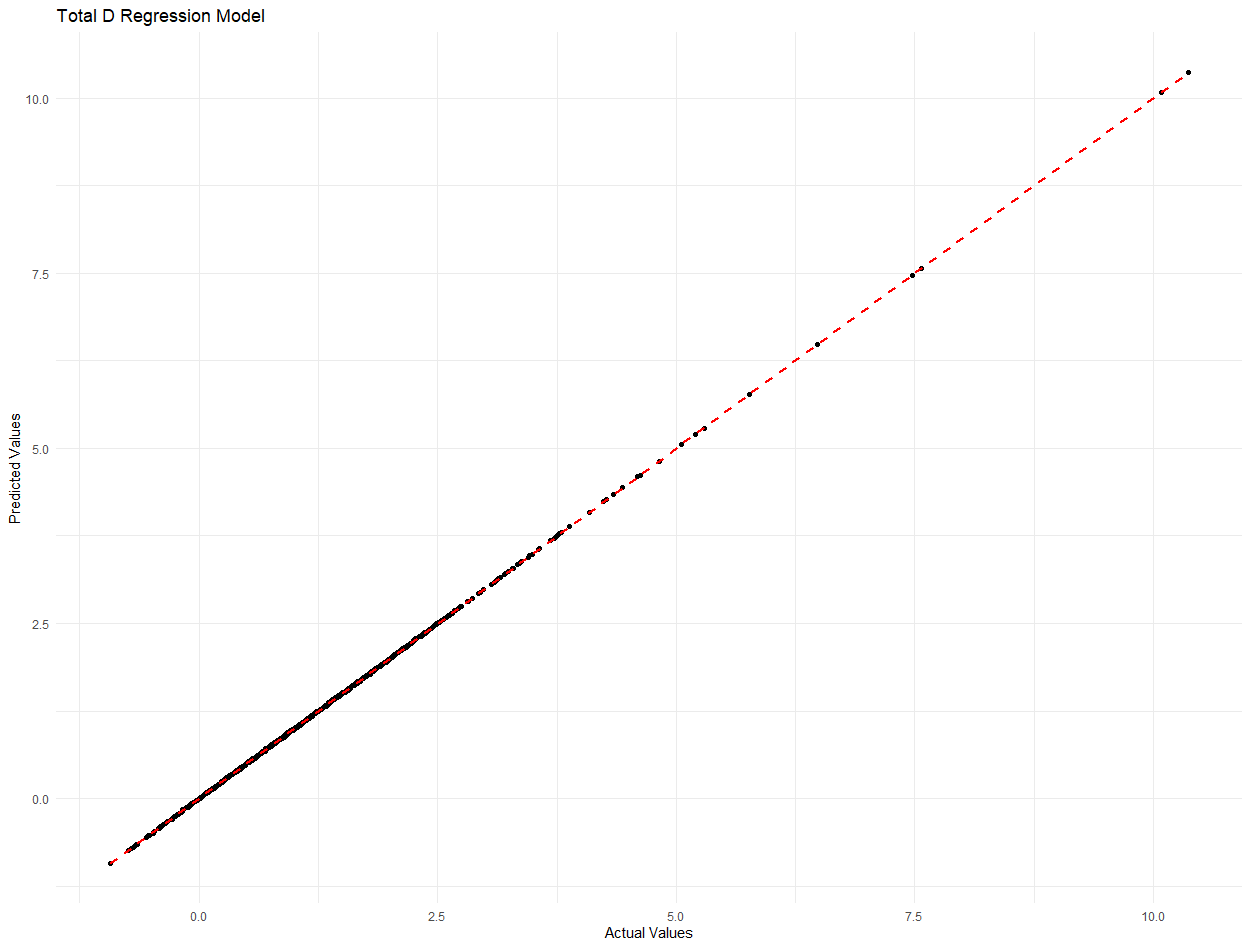}
\caption{Predicated vrs actual values of the $D_{\text{final}}$. The red dashed line illustrates the linear relationship between the predicted and actual values.}
\label{fig:image12}
\end{figure}

It's worth noting that the fitted values are the based solely on the relationship between the $D$ indices and $D_{\text{final}}$. Fitting \emph{any} random sample of exams will result in values that are close to these, regardless of the distribution of that sample. Even if every exam in the sample were abnormal, the value of $C$ would be $\approx$ 0.640.

More importantly, since the $D$ indices are z-score normalized and, by assumption, independent of each other, the mean of $D_{\text{final}}$ should equal the intercept of the regression equation. As modeled by BAD, the mean of $D_{\text{final}}$ is $\approx$ 0.640.

As demonstrated by the studies in our meta-analysis, however, the mean of $D_{\text{final}}$ \emph{for a sample} won't necessarily be $\approx$ 0.640 (see Table~\ref{tab:d_stats}). Using the same fitted model we can demonstrate that, when the means of the $D$ indices are non-zero, they shift $D_{\text{final}}$ in left or right direction, based on the signs of their means (±) and magnitudes of their coefficients. Anytime the mean of $D_{\text{final}}$ for a sample is not $C$ ($\approx$ 0.640), it's because the means of the $D$ indices are non-zero.

Perhaps not surprisingly, the absolute percentage deviation between the modes in our sample and estimated means for source measures are greatest for $D_{\text{aa}}$, $D_{\text{am}}$ , and $D_{\text{p}}$ (aside from $D_{\text{e}}$, whose deviation is large primarily because the absolute values involved are so small).

\begin{longtable}[]{@{}
  >{\raggedright\arraybackslash}p{(\linewidth - 10\tabcolsep) * \real{0.1361}}
  >{\raggedleft\arraybackslash}p{(\linewidth - 10\tabcolsep) * \real{0.0948}}
  >{\raggedright\arraybackslash}p{(\linewidth - 10\tabcolsep) * \real{0.4184}}
  >{\raggedleft\arraybackslash}p{(\linewidth - 10\tabcolsep) * \real{0.0948}}
  >{\raggedleft\arraybackslash}p{(\linewidth - 10\tabcolsep) * \real{0.0890}}
  >{\raggedleft\arraybackslash}p{(\linewidth - 10\tabcolsep) * \real{0.1669}}@{}}
\caption{Percent deviation between the empirical mode and estimted means for the source measures of the $D$ indices.}
\label{tab:pct_deviation}
\tabularnewline
\toprule\noalign{}
\begin{minipage}[b]{\linewidth}\centering
\textbf{Index}
\end{minipage} & \begin{minipage}[b]{\linewidth}\centering
\textbf{Mode}
\end{minipage} & \begin{minipage}[b]{\linewidth}\centering
\textbf{Source Measure}
\end{minipage} & \begin{minipage}[b]{\linewidth}\centering
\textbf{Mode}
\end{minipage} & \begin{minipage}[b]{\linewidth}\centering
\textbf{$\beta_{0}$}
\end{minipage} & \begin{minipage}[b]{\linewidth}\centering
\textbf{\% deviation}
\end{minipage} \\
\midrule\noalign{}
\endfirsthead
\toprule\noalign{}
\begin{minipage}[b]{\linewidth}\centering
\textbf{Index}
\end{minipage} & \begin{minipage}[b]{\linewidth}\centering
\textbf{Mode}
\end{minipage} & \begin{minipage}[b]{\linewidth}\centering
\textbf{Source Measure}
\end{minipage} & \begin{minipage}[b]{\linewidth}\centering
\textbf{Mode}
\end{minipage} & \begin{minipage}[b]{\linewidth}\centering
\textbf{$\beta_{0}$}
\end{minipage} & \begin{minipage}[b]{\linewidth}\centering
\textbf{\% deviation}
\end{minipage} \\
\midrule\noalign{}
\endhead
\bottomrule\noalign{}
\endlastfoot
$D_{\text{aa}}$ & 0.78 & art\_avg ($\mu$m) & 511 & 614 & 20\% \\
$D_{\text{am}}$ & 0.89 & art\_max ($\mu$m) & 390 & 488 & 25\% \\
$D_{\text{b}}$ & -0.16 & change\_back (mm) & 0.04 & 0.04 & 0\% \\
$D_{\text{e}}$ & 0.07 & ele\_b\_bfs\_8mm\_thinnest ($\mu$m) & 5 & 4 & 20\% \\
$D_{\text{f}}$ & 0.26 & change\_front (mm) & 0.03 & 0.03 & 0\% \\
$D_{\text{k}}$ & -0.21 & k\_max\_front\_d (D) & 44.7 & 45.0 & 1\% \\
$D_{\text{p}}$ & 0.76 & rpi\_avg & 1.02 & 0.90 & 12\% \\
$D_{\text{t}}$ & -0.27 & pachy\_min ($\mu$m) & 529 & 540 & 2\% \\
$D_{\text{y}}$ & -0.05 & pachy\_min\_y (mm) & -0.22 & -0.24 & 9\% \\
\end{longtable}

$D_{\text{aa}}$, $D_{\text{am}}$ , and
$D_{\text{p}}$ are the likely reason that
$D_{\text{final}}$ varies from $C$ in the published
studies.

\subsection{Variance Inflation Factor}

Variance Inflation Factor (VIF) is a metric used to assess multicollinearity in a multiple regression model. VIF quantifies this by measuring how much the variance of a regression coefficient is inflated due to multicollinearity. Specifically, the VIF for each predictor variable is calculated as the reciprocal of $1 - R^2$, where $R^2$ represents the coefficient of determination from regressing that predictor on all other predictors in the model. A VIF value of 1 indicates no multicollinearity, while values greater than 5 or 10, depending on the context, suggest that the predictor variable is redundant in the presence of others, which can make the model\textquotesingle s coefficients unstable and their interpretation unreliable. The VIF values $D_{\text{aa}}$, $D_{\text{am}}$ , and $D_{\text{p}}$ are all greater than 5.

\begin{longtable}[]{@{}
  >{\raggedleft\arraybackslash}p{(\linewidth - 16\tabcolsep) * \real{0.1283}}
  >{\raggedleft\arraybackslash}p{(\linewidth - 16\tabcolsep) * \real{0.1111}}
  >{\raggedleft\arraybackslash}p{(\linewidth - 16\tabcolsep) * \real{0.1111}}
  >{\raggedleft\arraybackslash}p{(\linewidth - 16\tabcolsep) * \real{0.0938}}
  >{\raggedleft\arraybackslash}p{(\linewidth - 16\tabcolsep) * \real{0.1111}}
  >{\raggedleft\arraybackslash}p{(\linewidth - 16\tabcolsep) * \real{0.1111}}
  >{\raggedleft\arraybackslash}p{(\linewidth - 16\tabcolsep) * \real{0.1111}}
  >{\raggedleft\arraybackslash}p{(\linewidth - 16\tabcolsep) * \real{0.1111}}
  >{\raggedleft\arraybackslash}p{(\linewidth - 16\tabcolsep) * \real{0.1111}}@{}}
\caption{Variance Inflation Factor (VIF) of the $D$ indices.}
\label{tab:vif}
\tabularnewline
\toprule\noalign{}
\begin{minipage}[b]{\linewidth}\raggedleft
\emph{\textbf{D\textsubscript{aa}}}
\end{minipage} & \begin{minipage}[b]{\linewidth}\raggedleft
\emph{\textbf{D\textsubscript{am}}}
\end{minipage} & \begin{minipage}[b]{\linewidth}\raggedleft
\emph{\textbf{D\textsubscript{b}}}
\end{minipage} & \begin{minipage}[b]{\linewidth}\raggedleft
\emph{\textbf{D\textsubscript{e}}}
\end{minipage} & \begin{minipage}[b]{\linewidth}\raggedleft
\emph{\textbf{D\textsubscript{f}}}
\end{minipage} & \begin{minipage}[b]{\linewidth}\raggedleft
\emph{\textbf{D\textsubscript{k}}}
\end{minipage} & \begin{minipage}[b]{\linewidth}\raggedleft
\emph{\textbf{D\textsubscript{p}}}
\end{minipage} & \begin{minipage}[b]{\linewidth}\raggedleft
\emph{\textbf{D\textsubscript{t}}}
\end{minipage} & \begin{minipage}[b]{\linewidth}\raggedleft
\emph{\textbf{D\textsubscript{y}}}
\end{minipage} \\
\midrule\noalign{}
\endfirsthead
\toprule\noalign{}
\begin{minipage}[b]{\linewidth}\raggedleft
\emph{\textbf{D\textsubscript{aa}}}
\end{minipage} & \begin{minipage}[b]{\linewidth}\raggedleft
\emph{\textbf{D\textsubscript{am}}}
\end{minipage} & \begin{minipage}[b]{\linewidth}\raggedleft
\emph{\textbf{D\textsubscript{b}}}
\end{minipage} & \begin{minipage}[b]{\linewidth}\raggedleft
\emph{\textbf{D\textsubscript{e}}}
\end{minipage} & \begin{minipage}[b]{\linewidth}\raggedleft
\emph{\textbf{D\textsubscript{f}}}
\end{minipage} & \begin{minipage}[b]{\linewidth}\raggedleft
\emph{\textbf{D\textsubscript{k}}}
\end{minipage} & \begin{minipage}[b]{\linewidth}\raggedleft
\emph{\textbf{D\textsubscript{p}}}
\end{minipage} & \begin{minipage}[b]{\linewidth}\raggedleft
\emph{\textbf{D\textsubscript{t}}}
\end{minipage} & \begin{minipage}[b]{\linewidth}\raggedleft
\emph{\textbf{D\textsubscript{y}}}
\end{minipage} \\
\midrule\noalign{}
\endhead
\bottomrule\noalign{}
\endlastfoot
14.182 & 9.762 & 2.933 & 3.11 & 1.651 & 1.215 & 6.542 & 2.607 & 1.154 \\
\end{longtable}

\section{Logistic Regression}

Our regression model is able to replicate BAD because it was fitted using the output variable ($D_{\text{final}}$) and the predictor variables (the $D$ indices) of every exam in our sample. Using linear regression, however, we only replicate BAD.

The Belin/Ambrósio Deviation model works differently. It is designed to distinguish between normal and keratoconic corneas, which is typically accomplished through \emph{logistic} regression.

The output of a logistic regression model is a probability value between 0 and 1, representing the likelihood that a given observation belongs to a specific group. An optimal cutoff between groups is typically chosen to maximize the Youden Index (sensitivity + specificity - 1) on a Receiver Operating Characteristic (ROC) curve, striking an optimal balance between true positive and false positive rates. This is the more likely reason that $D_{\text{final}}$ has a different threshold for abnormal ($\ge$ 3.0 SD) than the $D$ indices ($\ge$ 2.6 SD), which use an arbitrary number of standard deviations from the mean as their cutoff for abnormal.

Of course, $D_{\text{final}}$ is measured in standard deviations from the mean, not percentage probability. Through the application of the link function, though---typically the logit (which is the inverse of the logistic, hence ``logistic'' regression)---probabilities can be transformed back to the model's original units of measure. The logit enables users of the model to interpret the model's results on the same scale as the input variables.

The logit is defined as:

\begin{equation}
x = \text{logit}(p) = \ln(\frac{p}{1 - p})
\label{eq:logit}
\end{equation}

where:

\begin{itemize}
\item
  $p$ is a probability,
\item
  $x$ is a real-valued number.
\end{itemize}

If we presume that $D_{\text{final}}$ is the logit of the output of a logistic regression equation, we can convert $D_{\text{final}}$ back to a probability by applying the logistic:

\begin{equation}
p = \text{logistic}(x) = \frac{1}{1 + \ e^{- x}}\ 
\label{eq:logistic}
\end{equation}

If we calculate the logistic of $C$, the result tells us that when all of the $D$ indices are at the mean values, an exam has a 65.5\% probability of being keratoconic.

\begin{figure}
\centering
\includegraphics[width=\textwidth]{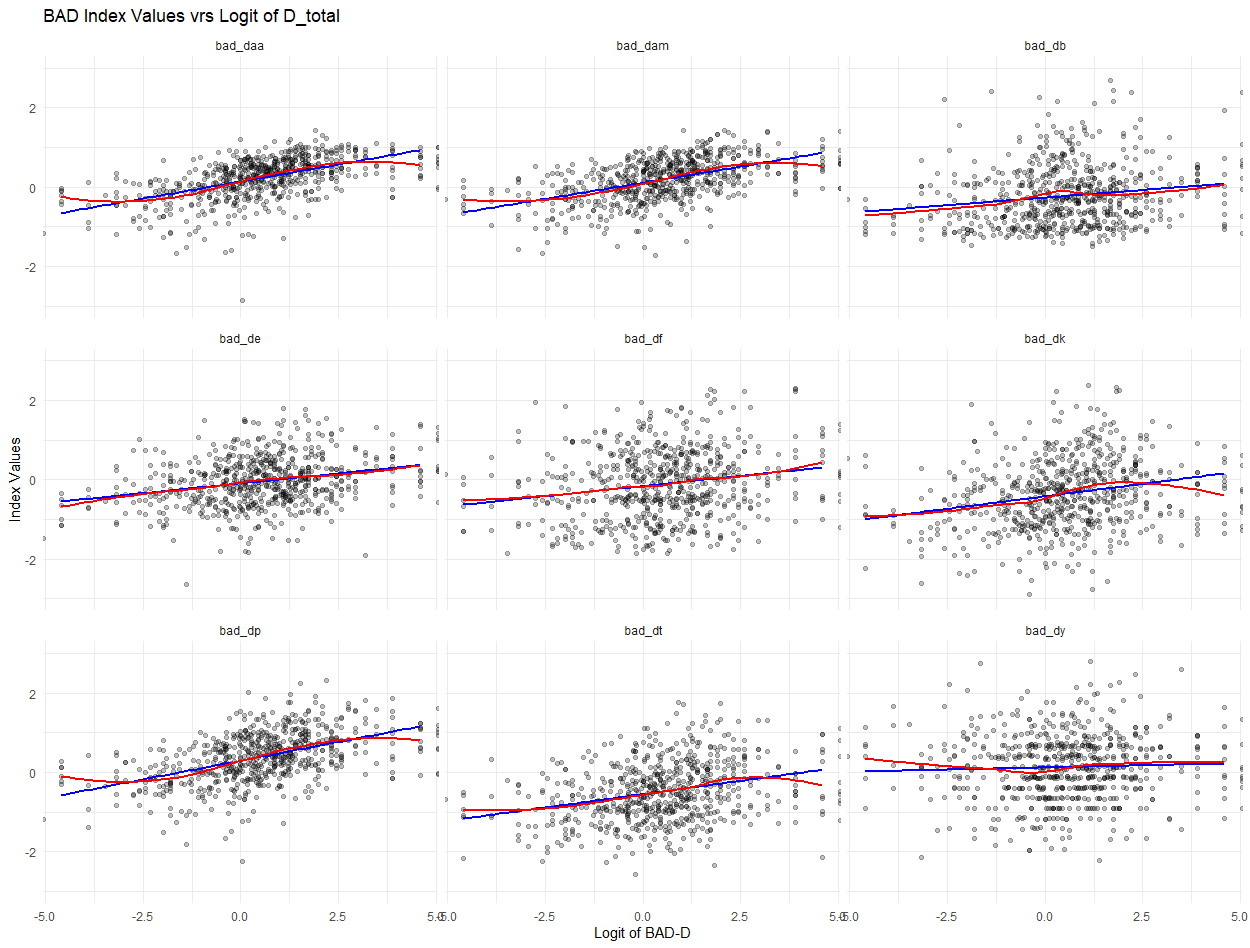}
\caption{BAD Index values vs Logit of $D_{\text{final}}$.}
\label{fig:image13}
\end{figure}

For logistic regression, it is assumed that the predictors have a linear relationship with the logit of the response variable. Figure~\ref{fig:image13} contains scatter plots of the model's $D$ indices against the logit of $D_{\text{final}}$. The blue line shows the linear relationship between the two values, and the red line shows a fitted LOESS curve. Any difference between the blue and red lines represents a deviation from linearity. As before, $D_{\text{aa}}$, $D_{\text{am}}$, and $D_{\text{p}}$---and potentially $D_{\text{t}}$---have all exhibit curvilinear relationships near extreme values of the response.

Moreover, the regression lines for $D_{\text{y}}$ exhibit are nearly flat, suggesting the parameter contributes very little to the model.

Ultimately, though, we cannot confirm that BAD was built using logistic regression. We do not have access to the exams that comprised the original normative database, so we cannot validate our logistic regression model in the same way that we validated our linear regression model. As such, we cannot validate that the logistic of $C$ gives us the probability of an exam being keratoconic when all of the $D$ indices are zero, much less that the probability is 65.5\%.

\section{Discussion}

Despite its clinical effectiveness in distinguishing between normal and abnormal corneas, the Belin/Ambrósio Deviation (BAD) model presents a number of interpretive challenges.

Studies have pointed out that it can be confusing when one or more the $D$ indices that appear on the ``Enhanced Ectasia Display'' is suspicious or abnormal, but $D_{\text{final}}$ itself is normal \cite{song2023, belin2015}. The apparent lack of agreement between the $D$ indices and $D_{\text{final}}$ can make it hard for clinicians to understand the relationship between the two.

\begin{figure}
\centering
\includegraphics[width=\textwidth]{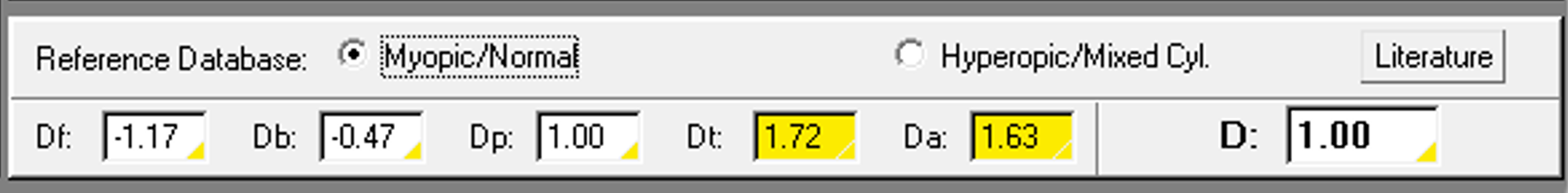}
\caption{Screenshot showing suspicious $D$ indices but normal $D_{\text{final}}$.}
\end{figure}

It can be even more confusing when the five $D$ indices are all normal, but $D_{\text{final}}$ is suspicious or abnormal.

\begin{figure}
\centering
\includegraphics[width=\textwidth]{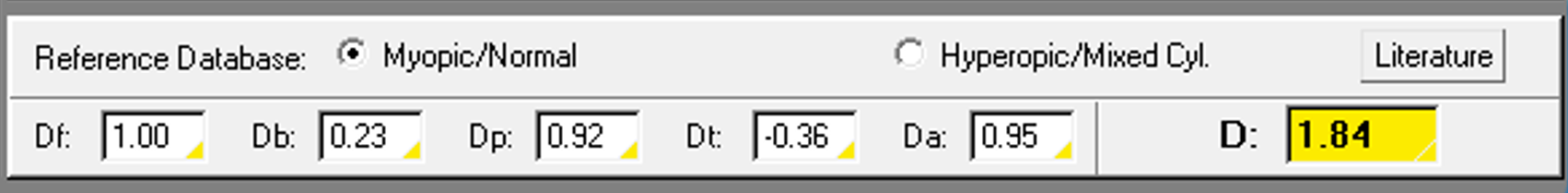}
\caption{Screenshot showing normal $D$ indices but suspicious $D_{\text{final}}$.}
\end{figure}

The only explanation that the model offers are the five index values shown on the ``Enhanced Ectasia Display''. Not realizing that $D_{\text{final}}$ is actually based \emph{nine} indices---and not being able to see the other four---can make it especially difficult to identify the underlying reasons when an exam is categorized as suspicious or abnormal.

Typically, the mathematical nature of regression models allows them to provide explanations for their results. By quantifying the contributions of individual predictors, regression models effectively ``explain'' the effect that each variable has on the outcome. Unfortunately, when a model is built using highly correlated predictors---like $D_{\text{aa}}$, $D_{\text{am}}$, and $D_{\text{p}}$---it cannot differentiate their individual effects, so it assigns arbitrary weights to each correlated predictor. While these weights accurately reflect the combined effect of the correlated predictors, they fail to reliably indicate the unique contribution of each predictor, making interpretation challenging.

The curvilinear relationships observed between some of the $D$ indices and $D_{\text{final}}$ further complicate interpretation of the model. The presence of non-linear patterns suggests that a simple linear regression approach may be insufficient to capture the complexity of the relationships among the indices. This limitation could lead to inaccuracies in the classification of corneas, particularly for cases that lie near the thresholds of normality and abnormality. Incorporating non-linear modeling techniques or re-evaluating the transformation of predictor variables could improve the model\textquotesingle s accuracy and its ability to differentiate between normal and ectatic corneas.

The fact that BAD cannot explain itself makes $D_{\text{final}}$ much more difficult to interpret and understand, particularly in terms of physical characteristics of the cornea.

The $D$ indices, though, are z-score normalized variables, representing how far a given measurement deviates from the mean of a
normative population, in units of standard deviations. This makes them relatively easy to interpret in terms of physical characteristics of the cornea---each $D$ index directly relates to a specific corneal measurement, such as thickness or elevation, allowing clinicians to
understand whether a value is within the normal range or indicative of an abnormality.

It\textquotesingle s reasonable to presume that $D_{\text{final}}$ should be equally easy to understand: It\textquotesingle s simply the weighted sum of the $D$ indices. But $D_{\text{final}}$ cannot be interpreted as a z-score. Because each of the $D$ indices is based on its own statistical properties, $D_{\text{final}}$ is an abstract measure, representing the cumulative distance from the mean across multiple dimensions.

Expectng $D_{\text{final}}$ to behave the same as the $D$ indices is one of the reasons that its non-zero mean seems paradoxical. We expect it to be zero, but---as multiple studies have shown---$D_{\text{final}}$ has a non-zero mean. Unfortunately, none of the studies in our meta-analysis have drawn attention to that fact, much less explored why, leaving the paradox unexplained.

The constant term ($C$) in the regression equation is one reason that $D_{\text{final}}$ has a non-zero mean. Even when all the $D$ indices are at their normative values, $D_{\text{final}}$ remains elevated at approximately 0.64, indicating a baseline shift. This suggests that BAD includes a baseline bias that elevates the $D_{\text{final}}$ score independently of the corneal characteristics being measured. This makes sense if we presume---as is likely the case---that BAD captures only a subset of the corneal features that are characteristic of keratoconus. Not knowing for sure, however, complicates clinical interpretation.

Many of the same studies have shown that $D_{\text{aa}}$, $D_{\text{am}}$, and $D_{\text{p}}$ also have non-zero means, which is another reason that $D_{\text{final}}$ has a non-zero mean. It's important to note, however, that, as z-scores, the means of $D_{\text{aa}}$, $D_{\text{am}}$, and $D_{\text{p}}$ \emph{should} all be zero, raising important questions about the normative database(s) used to develop BAD. One plausible explanation is that the normative data used to define the model may not accurately represent the diverse populations encountered in clinical practice. This discrepancy suggests a need to revisit and potentially update the normative datasets or recalibrate the model to better align with real-world patient characteristics.

One challenge, of course, is that BAD was trained on unpublished data, which makes validation and/or recalibration difficult. BAD---and the industry more broadly---could benefit from published datasets that could be used to validate (or enhance) existing models and enable further research. Ideally, these would be large enough and diverse enough to address other documented challenges with BAD, particularly related to categorical variables such as age, ethnicity, and others \cite{oculus2022}.

In conclusion, while the BAD model remains a valuable tool for screening and diagnosing ectatic corneal diseases, this study highlights several areas where its design and application could be improved. Addressing these limitations could lead to a more accurate and reliable model, ultimately improving patient outcomes and the management of corneal ectasia.

\bibliographystyle{plain} 
\bibliography{paradox.bib}

\begin{thebibliography}{10}

\bibitem{ambrosio2011b}
R.~Ambrósio, A.~L.~C. Caiado, F.~P. Guerra, et~al.
\newblock Novel pachymetric parameters based on corneal tomography for
  diagnosing keratoconus.
\newblock {\em Journal of Refractive Surgery}, 27(10):753--758, 2011.

\bibitem{ambrosio2013}
R.~Ambrósio, F.~Faria-Correia, I.~Ramos, B.~F. Valbon, B.~Lopes, D.~Jardim,
  and A.~Luz.
\newblock Enhanced screening for ectasia susceptibility among refractive
  candidates: The role of corneal tomography and biomechanics.
\newblock {\em Current Ophthalmology Reports}, 1(1):28--38, 2013.

\bibitem{ambrosio2017}
R.~Ambrósio, B.~T. Lopes, F.~Faria-Correia, M.~Q. Salomão, J.~Bühren, C.~J.
  Roberts, A.~Elsheikh, R.~Vinciguerra, and P.~Vinciguerra.
\newblock Integration of scheimpflug-based corneal tomography and biomechanical
  assessments for enhancing ectasia detection.
\newblock {\em Journal of Refractive Surgery}, 33(7):434--443, 2017.

\bibitem{ambrosio2011}
R.~Ambrósio, L.~P. Nogueira, D.~L. Caldas, B.~M. Fontes, A.~Luz, J.~O. Cazal,
  M.~R. Alves, and M.~W. Belin.
\newblock Evaluation of corneal shape and biomechanics before lasik.
\newblock {\em International Ophthalmology Clinics}, 51(2):11--38, 2011.

\bibitem{ambrosio2023}
Renato Ambrósio et~al.
\newblock Optimized artificial intelligence for enhanced ectasia detection
  using scheimpflug-based corneal tomography and biomechanical data.
\newblock {\em American Journal of Ophthalmology}, 251:126--142, 2023.

\bibitem{belin2013b}
M.~Belin and R.~Ambrósio.
\newblock Scheimpflug imaging for keratoconus and ectatic disease.
\newblock {\em Indian Journal of Ophthalmology}, 61(8):401, 2013.

\bibitem{belin2012}
M.~Belin, S.~Khachikian, and R.~Jr.
\newblock {\em Keratoconus / Ectasia Detection with a Modified (Enhanced)
  Reference Surface Belin / Ambrósio Enhanced Ectasia Display III}.
\newblock 2012.

\bibitem{belin2023}
M.~W. Belin.
\newblock {\em Scheimpflug Imaging for Keratoconus and Ectatic Disease}, pages
  203--220.
\newblock 2023.

\bibitem{belin2015}
M.~W. Belin, R.~Ambrósio~Jr, and A.~Steinmüller.
\newblock {\em Simplifying preoperative keratoconus screening}.
\newblock Oculus Optikgeräte GmbH, 3rd edition, 2015.

\bibitem{boyd2020}
B.~M. Boyd, J.~Bai, M.~Borgstrom, and M.~W. Belin.
\newblock Comparison of chinese and north american tomographic parameters and
  the implications for refractive surgery screening.
\newblock {\em Asia-Pacific Journal of Ophthalmology}, 9(2):117--125, 2020.

\bibitem{ding2024}
L.~Ding, L.~Niu, W.~Shi, X.~Zhou, and Y.~Qian.
\newblock Influence of corneal diameter on the accuracy of corneal tomography
  in patients with forme fruste keratoconus or thin corneas.
\newblock {\em Clinical and Experimental Optometry}, pages 1--7, 2024.

\bibitem{ding2020}
L.~Ding, J.~Wang, L.~Niu, W.~Shi, and Y.~Qian.
\newblock Pentacam scheimpflug tomography findings in chinese patients with
  different corneal diameters.
\newblock {\em Journal of Refractive Surgery}, 36(10):688--695, 2020.

\bibitem{oculus2018}
OCULUS~Optikgeräte GmbH.
\newblock {\em Pentacam user guide (Basic, HR, AXL)}.
\newblock OCULUS Optikgeräte GmbH, 2018.

\bibitem{oculus2022}
OCULUS~Optikgeräte GmbH.
\newblock Optimizing clinical use of the belin ambrósio (bad) display - prof.
  michael w. belin.
\newblock YouTube, 2022.

\bibitem{gomes2015b}
J.~A.~P. Gomes, D.~Tan, C.~J. Rapuano, M.~W. Belin, R.~Ambrósio, J.~L. Guell,
  F.~Malecaze, K.~Nishida, and V.~S. Sangwan.
\newblock Global consensus on keratoconus and ectatic diseases.
\newblock {\em Cornea}, 34(4):359--369, 2015.

\bibitem{hashemi2016}
H.~Hashemi, A.~Beiranvand, A.~Yekta, A.~Maleki, N.~Yazdani, and M.~Khabazkhoob.
\newblock Pentacam top indices for diagnosing subclinical and definite
  keratoconus.
\newblock {\em Journal of Current Ophthalmology}, 28(1):21--26, 2016.

\bibitem{lin2022}
Q.~Lin and Z.~Shen.
\newblock Effect of white-to-white corneal diameter on biomechanical indices
  assessed by pentacam scheimpflug corneal tomography and corneal visualization
  scheimpflug technology.
\newblock {\em International Ophthalmology}, 42(5):1537--1543, 2022.

\bibitem{matharu2022}
K.~S. Matharu, J.~Ma, Y.~Wang, and V.~Jhanji.
\newblock {\em Biomechanics of Keratoconus}, pages 23--29.
\newblock 2022.

\bibitem{ramos2012}
I.~Ramos, F.~F. Correia, B.~Lopes, M.~Q. Salomão, and R.~O. Correa.
\newblock Topometric and tomographic indices for the diagnosis of keratoconus.
\newblock {\em International Journal of Keratoconus and Ectatic Corneal
  Diseases}, 1(2):92--99, 2012.

\bibitem{shetty2017}
R.~Shetty, H.~Rao, P.~Khamar, K.~Sainani, K.~Vunnava, C.~Jayadev, and
  L.~Kaweri.
\newblock Keratoconus screening indices and their diagnostic ability to
  distinguish normal from ectatic corneas.
\newblock {\em American Journal of Ophthalmology}, 181:140--148, 2017.

\bibitem{song2023}
Y.~Song, Y.~Feng, M.~Qu, et~al.
\newblock Analysis of the diagnostic accuracy of belin/ambrósio enhanced
  ectasia and corvis st parameters for subclinical keratoconus.
\newblock {\em International Ophthalmology}, 43:1465--1475, 2023.

\bibitem{steinberg2015}
J.~Steinberg, S.~Aubke-Schultz, A.~Frings, et~al.
\newblock Correlation of the kisa\% index and scheimpflug tomography in
  ‘normal’, ‘subclinical’, ‘keratoconus-suspect’ and ‘clinically
  manifest’ keratoconus eyes.
\newblock {\em Acta Ophthalmologica}, 93(3), 2015.

\bibitem{toprak2023}
İ. Toprak, Ç. Martin, C.~E. Güneş, and J.~Alio.
\newblock Revisiting pentacam parameters in the diagnosis of subclinical and
  mild keratoconus based on different grading system definitions.
\newblock {\em Turkish Journal of Ophthalmology}, 53(6):324--335, 2023.

\bibitem{villavicencio2014}
O.~F. Villavicencio, F.~Gilani, M.~A. Henriquez, L.~Izquierdo, and R.~R.
  Ambrósio.
\newblock Independent population validation of the belin/ambrósio enhanced
  ectasia display: Implications for keratoconus studies and screening.
\newblock {\em International Journal of Keratoconus and Ectatic Corneal
  Diseases}, 3(1):1--8, 2014.

\end{thebibliography}

\end{document}